\documentclass[aps,prx, superscriptaddress,
 amsmath,amssymb, 
 reprint,%
]{revtex4-2}

\usepackage{graphicx}% Include figure files
\usepackage{subfigure}
\usepackage{dcolumn}% Align table columns on decimal point
\usepackage{multirow}
\usepackage{bm}% bold math
\usepackage{array}
\usepackage[colorlinks=true,linkcolor=blue,anchorcolor=blue,citecolor=blue,urlcolor=blue]{hyperref}
\usepackage{comment}
\usepackage[table,xcdraw]{xcolor}
\usepackage{float}
\usepackage{mathptmx}
\usepackage{amsmath}
\usepackage{etoolbox, xcolor}
\usepackage{soul}
\usepackage{amssymb}% http://ctan.org/pkg/amssymb
\usepackage{pifont}% http://ctan.org/pkg/pifont
\makeatletter
\def\ignorecitefornumbering#1{%
     \begingroup
         \@fileswfalse
         #1%                     % do \cite comand
    \endgroup
}
\makeatother

\begin{document}

\title{Computational Design of Two-Dimensional MoSi$_2$N$_4$ Family Field-Effect Transistor for \\Future \AA ngstr\"om-Scale CMOS Technology Nodes}

\author{Che Chen Tho}
\affiliation{Science, Mathematics and Technology Cluster, Singapore University of Technology and Design, Singapore 487372}

\author{Zongmeng Yang}
\affiliation{State Key Laboratory for Mesoscopic Physics and School of Physics, Peking University, Beijing 100871, China}

\author{Shibo Fang}
\affiliation{Science, Mathematics and Technology Cluster, Singapore University of Technology and Design, Singapore 487372}

\author{Shiying Guo}
\affiliation{College of Physics Science and Technology, Yangzhou University, Yangzhou 225002, China}

\author{Liemao Cao}
\affiliation{College of Physics and Electronic Engineering, Hengyang Normal University, Hengyang 421002, China}

\author{\\Chit Siong Lau}
\affiliation{Quantum Innovation Centre (Q. InC), Agency for Science Technology and Research (A*STAR), Singapore 138634}
\affiliation{Institute of Materials Research and Engineering (IMRE), Agency for Science Technology and Research (A*STAR), Singapore 138634}
\affiliation{Science, Mathematics and Technology Cluster, Singapore University of Technology and Design, Singapore 487372}

\author{Fei Liu}
\affiliation{School of Integrated Circuit and Beijing Advanced Innovation Center
for Integrated Circuits, Peking University, Beijing 100871, China}

\author{Shengli Zhang}
\affiliation{MIIT Key Laboratory of Advanced Display Materials and Devices, Jiangsu Engineering Research Center for Quantum Dot Display, Institute of Optoelectronics \& Nanomaterials, School of Materials Science and Engineering, Nanjing University of Science and Technology, Nanjing 210094, China}

\author{Jing Lu}
\affiliation{State Key Laboratory for Mesoscopic Physics and School of Physics, Peking University, Beijing 100871, China}

%\author{Kah-Wee Ang}
%\affiliation{Department of Electrical \& Computer Engineering, National University of Singapore, Singapore 117583, Singapore}

\author{L. K. Ang}
\email{ricky\_ang@sutd.edu.sg}
\affiliation{Science, Mathematics and Technology Cluster, Singapore University of Technology and Design, Singapore 487372}

\author{Lain-Jong Li}
\email{lanceli@nus.edu.sg}
\affiliation{Department of Materials Science \& Engineering, National University of Singapore 117575}

\author{Yee Sin Ang}
\email{yeesin\_ang@sutd.edu.sg}
\affiliation{Science, Mathematics and Technology Cluster, Singapore University of Technology and Design, Singapore 487372}

\begin{abstract}

Advancing complementary metal–oxide–semiconductor (CMOS) technology into the sub-1-nm \AA ngstr\"om-scale technology nodes is expected to involve alternative semiconductor channel materials, as silicon transistors encounter severe performance degradation at physical gate lengths below 10 nm. Two-dimensional (2D) semiconductors have emerged as strong candidates for overcoming short-channel effects due to their atomically thin bodies, which inherently suppress electrostatic leakage and improve gate control in aggressively scaled field-effect transistors (FETs). Among the growing library of 2D materials, the MoSi$_2$N$_4$ family -- a synthetic septuple-layered materials -- has attracted increasing attention for its remarkable ambient stability, suitable bandgaps, and favorable carrier transport characteristics, making it a promising platform for next-generation transistors. While experimental realization of sub-10-nm 2D FETs remains technologically demanding, computational device simulation using first-principles density functional theory combined with nonequilibrium Green’s function transport simulations provide a powerful and cost-effective route for exploring the performance limits and optimal design of ultrascaled FET. This review consolidates the current progress in the computational design of MoSi$_2$N$_4$ family FETs. We review the physical properties of  MoSi$_2$N$_4$ that makes them compelling candidates for transistor applications, as well as the simulated device performance and optimization strategy of  MoSi$_2$N$_4$ family FETs. Finally, we identify key challenges and research gaps, and outline future directions that could accelerate the practical deployment of MoSi$_2$N$_4$ family FET in the \AA ngstr\"om-scale CMOS era.

\end{abstract}

\maketitle

\section{\label{sec: introduction}Introduction}

Modern electronics demands aggressive miniaturization of transistors to support advanced digital technologies such as artificial intelligence (AI), telecommunication networks, and the Internet of Things. As device dimensions shrink, designing high-performance field-effect transistors (FETs) becomes increasingly challenging due to the physical limitations of silicon -- the foundational material of modern FETs. At physical gate lengths below 10 nm, silicon-based FETs suffer from severe mobility degradation and short-channel effects (SCE), which significantly degrade their performance. To sustain Moore’s law, complex device architectures such as FinFET, gate-all-around (GAA), and complementary FET (CFET) have been introduced to extend the scalability of silicon transistors into the nanometer regime \cite{zhang2024_Si_advanced_FET,duan2024_Si_finfet_to_gaafet,dutta2025_Si_mosfet_to_cfets}. Nonetheless, pushing device scaling beyond the 1-nm technology node, into the \AA ngstr\"om-scale era, necessitates alternative channel materials and radically new device concepts that transcend the conventional silicon paradigm.

\begin{figure*}
\centering
\includegraphics[width=\textwidth]{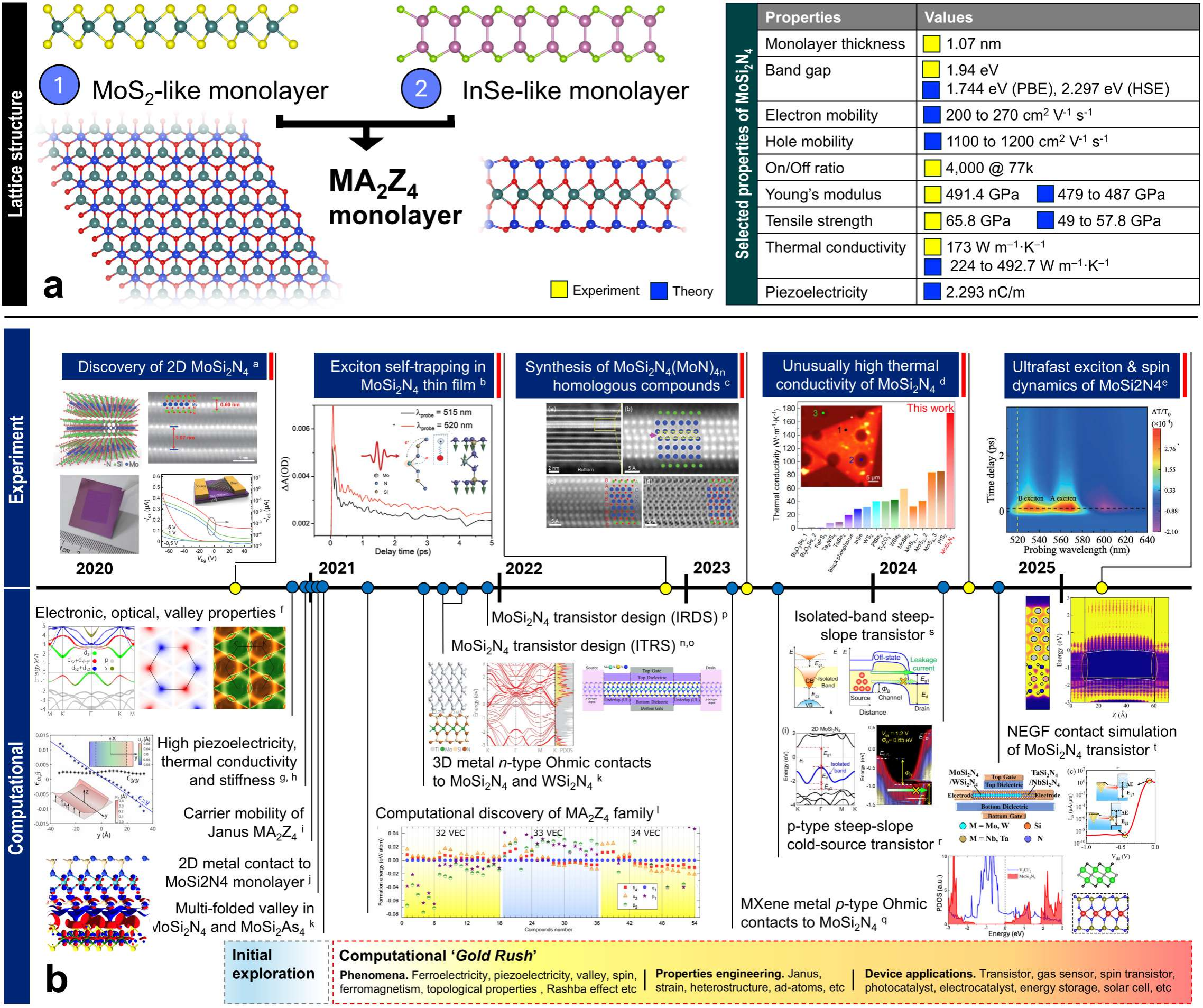}
\caption{\label{Fig1}\textbf{Overview of MoSi$_2$N$_4$ and the MA$_2$Z$_4$ monolayer family.} (a) Intercalation morphology of MA$_2$Z$_4$ monolayer and selected representative physical properties of MoSi$_2$N$_4$. (b) Current status of MA$_2$Z$_4$ monolayer for electronics applications. Images in (b) are extracted from Ref. a \cite{10.1063/1.4977487} Copyright 2022, American Association for the Advancement of Science; Ref. b \cite{huang2023_MoSi2N4_exciton_trapping} Copyright 2024, American Physical Society; Ref. c \cite{NSR} Copyright 2022, The Author(s); Ref. d \cite{he2024_MoSi2N4_thermal_conductivity} Copyright 2024, The Author(s); Ref. e \cite{wu2025_MoSi2N4_exciton_spindynamics} Copyright 2025, The Author(s); Ref. f \cite{Li2020} Copyright 2020, American Physical Society; Ref. g \cite{Guo2020} Copyright 2020, EPLA; Ref. h \cite{Mortazavi2021} Copyright 2020, Elsevier Ltd.; Ref. i \cite{guo2021_Rashba} Copyright 2021, The Royal Society of Chemistry; Ref. j \cite{Cao2021} Copyright 2021, The Author(s); Ref. k \cite{Yang2021} Copyright 2021, American Physical Society; Ref. l \cite{Wang2021_MA2Z4} Copyright 2021, The Author(s); Ref. m \cite{Wang2021} Copyright 2021, The Author(s);  Ref. n \cite{Sun2021_transistor} Copyright 2021, Royal Society of Chemistry, Ref. o \cite{Huang2021_transistor} Copyright 2021, American Physical Society; Ref. p \cite{Nandan2021_transistor} Copyright 2022, IEEE; Ref. q \cite{He2023} Copyright 2023, Royal Society of Chemistry; Ref. r \cite{qu2023_MoSi2N4FET} Copyright 2023, IEEE; Ref. s \cite{qu2024_FET_ISB} Copyright 2024, Elsevier B.V. and Science China Press; Ref. t \cite{Ying2024_MoSi2N4_FET} Copyright 2024, American Chemical Society.} 
\end{figure*}

Two-dimensional (2D) semiconductors such as transition metal dichalcogenides (TMDs) \cite{lee2012synthesis,Choi2017}, black phosphorus (BP) \cite{liu2015semiconducting} and indium selenide (InSe) \cite{Padilha2017} have emerged as promising channel materials for ultrascaled FETs at the \AA ngstr\"om-scale technology node. Owing to their atomically thin bodies and dangling-bond-free surfaces, 2D materials exhibit excellent electrostatic control without suffering from mobility degradation, even down to the monolayer limit \cite{Liu2019}. Furthermore, the van der Waals nature of 2D materials enables vertical stacking, making them inherently compatible with multistack nanosheet architectures \cite{lim2014_2Dstacking,guo2021_2Dstacking,hu2020_2Dstacking,qian2015_2Dstacking_modelling}, which are advantageous for high-performance and densely integrated transistor applications. Notably, the International Roadmap for Devices and Systems (IRDS) \cite{celano2024_2DmaterialIRDS,irisawa2024_2DmaterialIRDS} has formally recognized 2D semiconductors as key enablers for extending complementary metal–oxide–semiconductor (CMOS) technology into the \AA ngstr\"om-scale era—an evolution that cannot be readily achieved with conventional silicon transistors \cite{zheng2025continue}.

MoSi$_2$N$_4$, along with the broader family of MA$_2$Z$_4$ monolayers (where M is a transition metal from groups IVB, VB, or VIB; A = Si or Ge; Z = N, P, or As) \cite{Yin2022_Review_MoSi2N4, latychevskaia2024_Review_MoSi2N4, jin2024_Review_MoSi2N4}, has emerged as a promising 2D material family for electronics \cite{Wang2021_MA2Z4}, optoelectronics \cite{Liu2022, Cai2021_MoSe2, Cai2021_WSe2,Pham2021,Nguyen2022_C3N4,Wang2021_MoGe2N4, Guo2022_BP}, spintronics \cite{Ren2022}, energy harvesting \cite{He2022,Zeng2021,Xuefeng2022,Liu2022} and sensing \cite{xiao2022_gas, Bafekry2021_gas} applications. 
MA$_2$Z$_4$ monolayers are synthetic monolayers with an intercalated architectures composed of one MoS$_2$-like inner core layer sandwiched by InSe-like outer layers [Fig. \ref{Fig1}(a)]. Such monolayer has no known bulk counterparts, offering an entirely new structural framework for 2D material design.
The successful chemical vapor deposition (CVD) growth of high-quality MoSi$_2$N$_4$ and WSi$_2$N$_4$ semiconducting monolayers \cite{Hong2020}, as well as the homologous metallic sister structure -- MoSi$_2$N$_4$(MoN)$_{4n}$ ($n=1,2,3,\cdots$) \cite{NSR} -- has triggered a `computational gold rush' in which first-principles simulations are carried out intensively to unearth the physical properties of MoSi$_2$N$_4$ family. 
MoSi$_2$N$_4$, in particular, exhibits exceptional ambient stability, a suitable bandgap for transistor applications, high carrier mobilities surpassing those of MoS$_2$, and outstanding mechanical and thermal robustness. These attributes make MoSi$_2$N$_4$ highly attractive for next-generation electronic device applications.

Given the experimental challenges associated with fabricating and characterizing FETs with sub-10-nm physical gate length, computational modeling plays an essential role in guiding material selection and device design by predicting performance limits and identifying promising performance optimization strategies \cite{marin2018modeling,Quhe2021,meena2023_sub5nm_FET}. In this review, we summarize recent advances in the computational design of MoSi$_2$N$_4$ family FETs using the nonequilibrium Green’s function (NEGF) quantum transport formalism \cite{datta2002non}. This simulation approach -- often integrated with density functional theory (DFT) to provide accurate electronic structures -- enables predictive assessments of device performance, scaling behavior, and optimal geometry in the ballistic or near-ballistic quantum transport regime. We begin by reviewing the electronic properties and contact physics of MoSi$_2$N$_4$, followed by an introduction to key performance metrics relevant to ultrascaled FETs. Subsequently, we present a comprehensive survey of NEGF-based studies on MoSi$_2$N$_4$ family FETs. Finally, we discuss the current challenges and future opportunities in realizing MoSi$_2$N$_4$ transistors. By providing an up-to-date account on the computational-driven developments of MoSi$_2$N$_4$ family FETs, this review aims to chart a forward path for MoSi$_2$N$_4$ family in ultrascaling CMOS technology nodes into the \AA ngstr\"om-scale device era.

\section{\label{sec: MA2Z4} Background and Rationale of MoSi$_2$N$_4$ Family for Transistor Applications}

\begin{figure*}
\centering
\includegraphics[width=\textwidth]{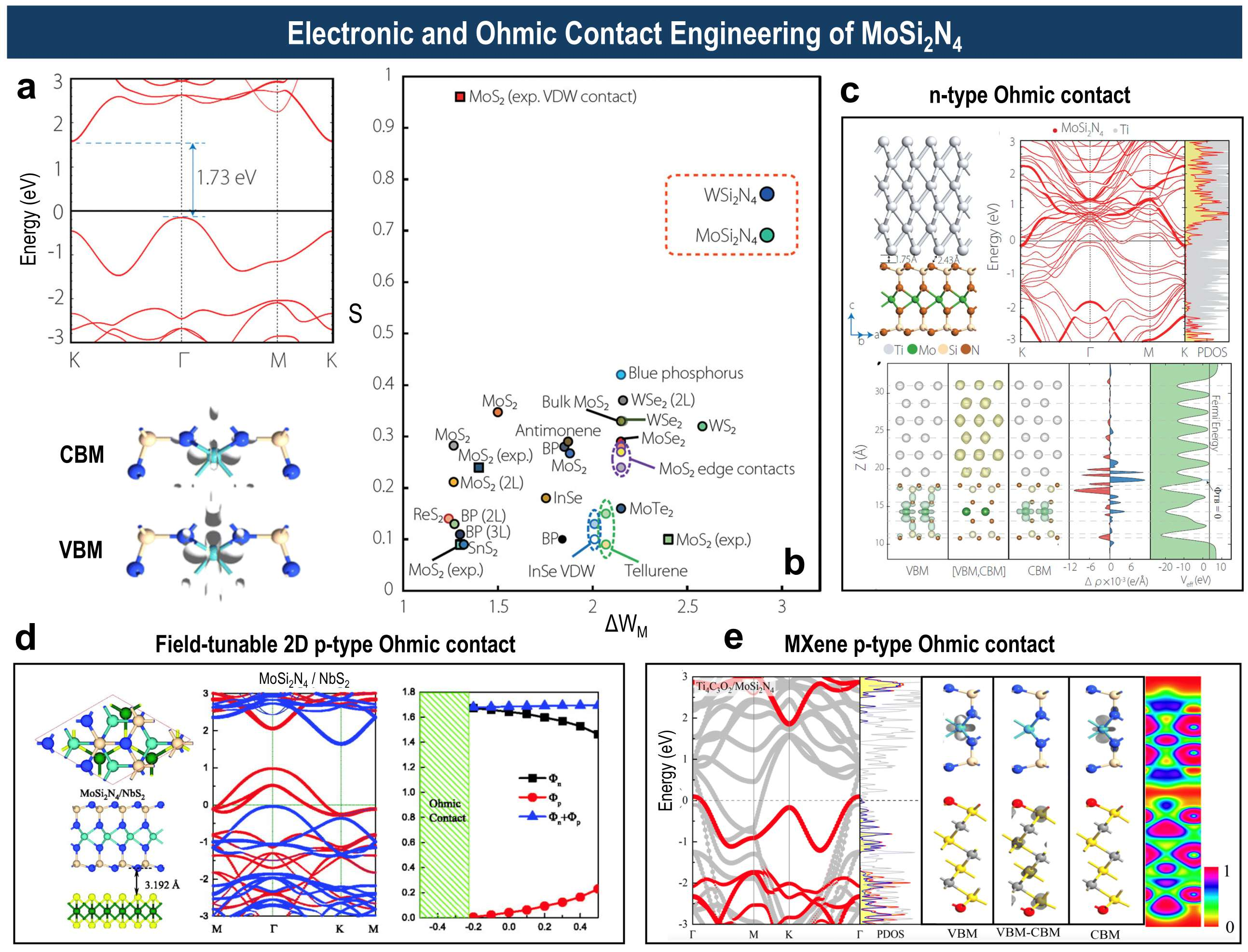}
\caption{\label{Fig2}\textbf{Electronic structures and contact properties of MoSi$_2$N$_4$.} (a) Band structures of MoSi$_2$N$_4$ \cite{Wang2021} and the charge distribution of CBM and VBM states \cite{He2023}. (b) Schottky-Mott $S$ parameter of MoSi$_2$N$_4$ and WSi$_2$N$_4$ exhibits one of the highest among other 2D semiconductors \cite{Wang2021}. (c) $n$-type Ohmic contact to MoSi$_2$N$_4$ using bulk metal \cite{Wang2021}. (d) Achieving gate-tunable $p$-type quasi-Ohmic or Ohmic contact to MoSi$_2$N$_4$ using 2D metal NbS$_2$ \cite{Cao2021}. (e) $p$-type Ohmic contact to MoSi$_2$N$_4$ using MXene \cite{He2023}. (a) Copyright 2021, The Author(s); Copyright 2023, Royal Society of Chemistry; (b), (c) Copyright 2021, The Author(s); (d) Copyright 2021, AIP Publishing; (e) Copyright 2023, Royal Society of Chemistry.}
\end{figure*}

Morphologically, MA$_2$Z$_4$ monolayers can be described as a structural intercalation of a MoS$_2$-like lattice with InSe-like lattice \cite{Wang2021_MA2Z4}. This septuple-layered configuration gives rise to four distinct phases of $\alpha$, $\beta$, $\gamma$, and $\delta$, leading to a diverse family of 2D monolayers, including semiconductors, metals, ferromagnets, topological insulators, and superconductors. MA$_2$Z$_4$ monolayers with 32 or 34 valence electrons are typically semiconducting, while those with 33 valence electrons may exhibit nonmagnetic metallic or ferromagnetic semiconducting characteristics \cite{Wang2021_MA2Z4}.

MoSi$_2$N$_4$, an experimentally synthesized member of MA$_2$Z$_4$ \cite{Hong2020}, exhibits excellent ambient stability and a suite of physical properties highly desirable for electronic applications. DFT calculations predict an indirect band gap of 1.73 eV [Fig. \ref{Fig2}(a)], which is in good agreement with experimental optical band gap measurements of approximately 1.94 eV. While the band gap is comparable to MoS$_2$, MoSi$_2$N$_4$ demonstrates significantly higher electron and hole mobilities [Fig. \ref{Fig1}(b)]. MoSi$_2$N$_4$ monolayer boasts a high Young's modulus of 491.4 GPa and tensile strength of 65.8 GPa, outperforming both MoS$_2$ and various MXenes. Additionally, MoSi$_2$N$_4$ exhibits one of the highest reported thermal conductivities among 2D materials \cite{Mortazavi2021}, which can be beneficial for heat dissipation in ultracompact electronics.

The experimental discovery of MoSi$_2$N$_4$ and WSi$_2$N$_4$ \cite{Hong2020}, and the high-throughput computational cataloging of the MA$_2$Z$_4$ monolayers \cite{Wang2021_MA2Z4} has sparked a computational “gold rush” into their properties and potential device applications. Based primarily on DFT simulations [see Fig. \ref{Fig1}(b) for a simplified timeline of MoSi$_2$N$_4$ studies related to electronics], MoSi$_2$N$_4$ family has been computationally proposed for a wide spectrum of applications, including transistors \cite{Zhao2021_WGe2N4, Sun2021_transistor, Huang2021_transistor, Ye2022_transistor, Nandan2021_transistor}, metal contacts \cite{Pham2021}, photodetectors \cite{Shu2022, Tho2022}, solar cells \cite{Guo2022_BP,Liu2022,Nguyen2022_C3N4,JinQuan2022,Bafekry2021}, photocatalysts \cite{Xuefeng2022,Zeng2021,He2022,Mortazavi2021}, light-emitting devices \cite{Cai2021_MoSe2,Cai2021_WSe2,JinQuan2022}, thermal management systems \cite{Mortazavi2021}, batteries \cite{Li2022_battery}, gas sensors \cite{xiao2022_gas, Bafekry2021_gas}, and piezoelectronics \cite{Zhong2021}. Despite the rapidly growing body of computational works, experimental progress remains relatively limited. Nevertheless, several landmark studies have demonstrated ultrafast spin dynamics \cite{wu2025_MoSi2N4_exciton_spindynamics}, strong excitonic effects~\cite{huang2023_MoSi2N4_exciton_trapping}, and unusually high thermal conductivity~\cite{Mortazavi2021, he2024_MoSi2N4_thermal_conductivity} in MoSi$_2$N$_4$. Furthermore, the recent synthesis of ultrathick metallic homologous compounds MoSi$_2$N$4$(MoN)${4n}$~\cite{NSR} significantly expands the material design space within this family. These developments underscore the largely untapped physical landscape of MoSi$_2$N$_4$ family and their potential as foundational materials for next-generation electronics.

\subsection{Contact and transport properties for transistor applications}

The metal contact plays a critical role in determining the performance of FET, particularly at the nanochannel regime. DFT simulations of metal contacts to MoSi$_2$N$_4$ and other MA$_2$Z$_4$ monolayers \cite{Ying2024_MoSi2N4_FET, Zhanhai2024_MoSi2N4_FET, Meng2022_Au, Wang2021, qu2023_MoSi2N4FET, shu2023_CrX2N4FET} have demonstrated the possibilities for achieving both $n$-type and $p$-type Ohmic contacts. 

Notably, the electronic states near the conduction band minimum (CBM) and valence band maximum (VBM) of MoSi$_2$N$_4$ are predominantly localized within the inner Mo–N sublayer and are effectively sandwiched between the outer Si–N sublayers [Fig. \ref{Fig2}(a)]. This unique spatial distribution of CBM and VBM states acts as a natural barrier that suppresses metal-induced gap states (MIGS) at the metal–semiconductor interface, even when an external metal is strongly bonded to MoSi$_2$N$_4$. As a result, MoSi$_2$N$_4$ exhibits a Schottky-Mott $S$ parameter higher than most other 2D semiconductors with significantly weakened Fermi-level pinning effect [Fig. \ref{Fig2}(b)]. Intriguingly, $n$-type Ohmic contacts with zero van der Waals (vdW) gap can be formed using bulk metals Ti and Sc [Fig. \ref{Fig2}(c)]. For $p$-type contact, DFT studies suggest that 2D metallic monolayer NbS$_2$ can form quasi-Ohmic contact with ultralow Schottky barrier height [Fig. \ref{Fig2}(d)] \cite{Cao2021}, while $p$-type `true' Ohmic contact to MoSi$_2$N$_4$, characterized by zero vdW barrier, has been demonstrated using MXene electrodes [Fig. \ref{Fig2}(e)] \cite{He2023}. Experimentally, long-channel ($L_\mathrm{G}$ = 30 $\mu$m) back-gated FET has been fabricated using MoSi2N4 single-crystal domains transferred onto SiO2/Si substrates \cite{Hong2020}. The device exhibits p-type behaviors (Fig. \ref{Fig2}(f)). An on/off ratio of $\sim$4000 has been achieved at 77 K and Ohmic contact behaviors was observed from the transfer characteristics (Fig. \ref{Fig2}(g) and (h)).

\begin{figure*}
\centering
\includegraphics[width=\textwidth]{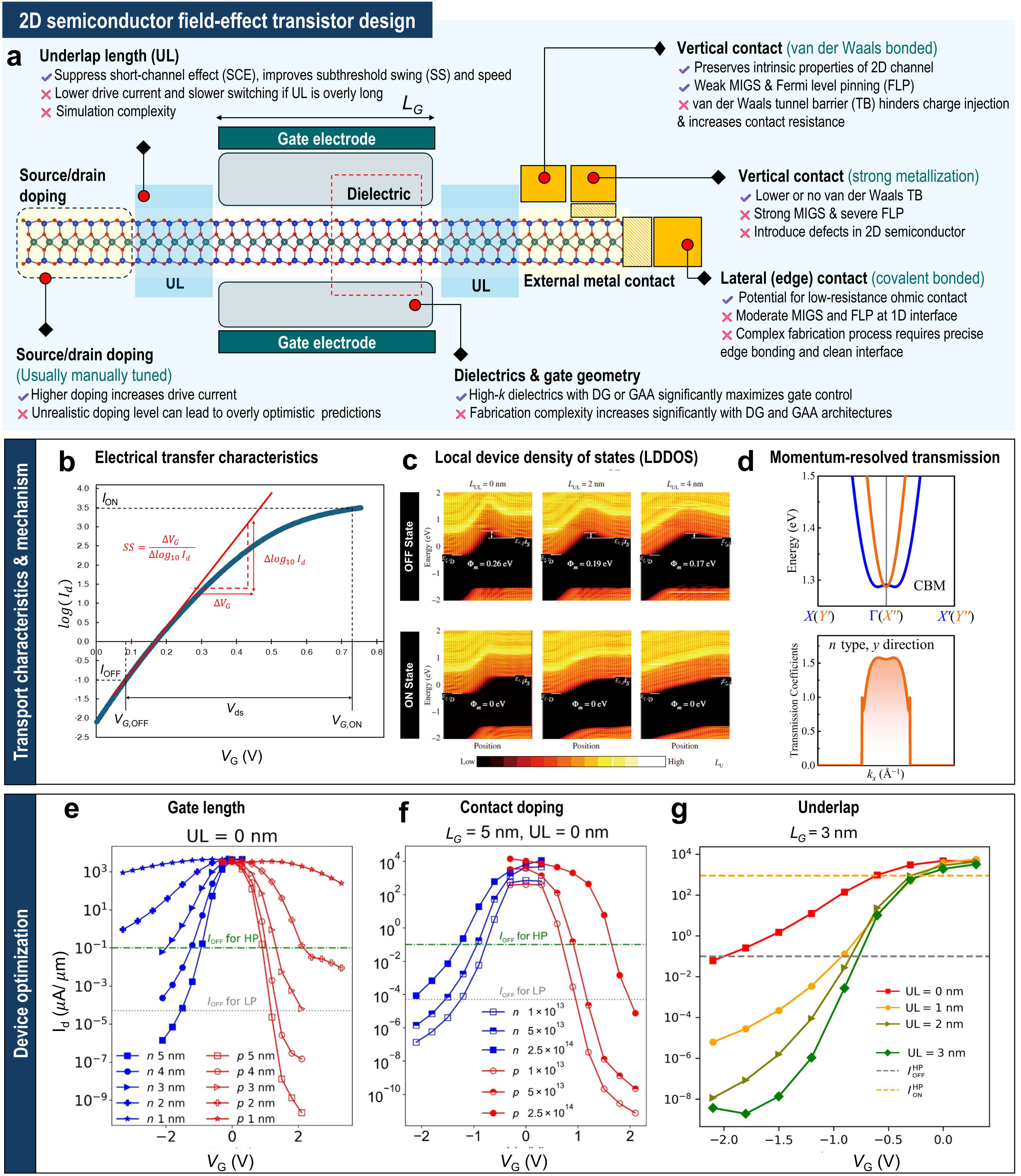}
\caption{\label{Fig3}\textbf{Computational design of 2D channel sub-10-nm FETs.} (a) Design considerations for sub-10-nm 2D FETs, highlighting trade-offs in underlap, gate, and contact strategies.
The transport characteristics can be assessed via (b) electrical transfer characteristics; (c) local device density of states (DDOS) \cite{Huang2021_transistor}; and (d) momentum-resolved transmission spectra. %The subthreshold swing ($SS$) can be obtained by taking the reciprocal of the steepest slope of the logarithmic drain current ($I_\mathrm{d}$) against applied gate voltage ($V_\mathrm{G}$), chosen around the FET OFF-state as shown in (b). 
(c) The LDDOS profiles of MoSi$_2$N$_4$ FET with different UL \cite{Huang2021_transistor}. The vanishing transport barrier at the ON-state, allowing a significant current to flow. $E_{F, D}$ and $E_{F, S}$ denotes the Fermi level at the drain and source, respectively. The transmission spectra (d) provides a microscopic view on the carrier transport physics of the device \cite{guo2022_TeO2}. Device performance can be optimized via gate length, contact doping and underlap. (e) Transfer characteristic of WSi$_2$N$_4$ FET at different $L_\mathrm{G}$, under a doping concentration of $5\times10^{13}$ cm$^{-2}$ \cite{li2023_WSi2N4}. (f) Transfer characteristic of the WSi$_2$N$_4$ FET under different \emph{n (p)}-type doping concentrations \cite{li2023_WSi2N4}. (g) Transfer characteristic of the WSi$_2$N$_4$ FET through different UL \cite{li2023_WSi2N4}. (e),(f),(g) Copyright 2023, American Physical Society; (c) and (d) Copyright 2021 and 2022, respectively, American Physical Society.}
\end{figure*}

\section{\label{sec: MA2Z4_FET}Computational Design of MoSi$_2$N$_4$ Family Transistors}

The computational simulations of sub-10-nm FETs are commonly benchmarked against semiconductor industry roadmaps. Historically, this is the International Technology Roadmap for Semiconductors (ITRS) \cite{ITRS}, which provided a bottom-up, primarily \textit{More Moore} driven forecast focused on the physical scaling of silicon transistors. The ITRS offered highly specific, quantitative targets for each technology node, including precise parameters like gate length ($L_\mathrm{G}$) and supply voltage ($V_\mathrm{ds}$).
Its successor, the International Roadmap for Devices and Systems (IRDS) \cite{IRDS_2020}, introduced in 2016, represents a significant shift. Recognizing the increasing limitations of traditional transistor scaling, the IRDS adopts a broader, top-down approach. IRDS expands its scope to encompass the entire electronics ecosystem, whereby alongside the trend of continued \textit{More Moore} scaling, there is additional emphasis on \textit{More than Moore} (e.g. heterogeneous integration), beyond CMOS (e.g. novel device physics), and comprehensive system-level considerations. As the physical scaling of silicon transistor slows down in pace and the concept of a universally defined 'node' has become less meaningful (since varying critical dimensions for the same node name are used across foundries), the IRDS no longer provides exact transistor parameters like $L_\mathrm{G}$ for future nodes. Instead, IRDS guides the semiconductor industry by outlining performance metrics and grand challenges for future technology nodes, focusing on system-level needs rather than prescriptive device dimensions.

Figure \ref{Fig3}(a) summarizes key design considerations in sub-10-nm 2D semiconductor FETs, highlighting the interplay between gate, contact engineering, and doping strategies. The underlap length (UL) -- an extra spacing between the gate edge and the source/drain regions -- effectively suppresses short-channel effect (SCE) and improves subthreshold swing (SS) by extending gate control. An optimal underlap of typically 1 to 4 nm in sub‑10-nm 2D FETs can significantly reduces gate and fringing capacitance \cite{Quhe2021}, thus improving the switching speed and energy efficiency. However, UL adds simulation complexity and an overly long UL could lead to higher channel resistance and reduce the overall device performance. Source/drain doping can enhance drive current but must be carefully and realistically optimized to avoid unrealistic predictions. Contact geometry plays a pivotal role in charge injection. van der Waals (vdW) vertical contact preserves the intrinsic properties of the contacted 2D semiconductor with weak Fermi level pinning (FLP), though they suffer from tunneling barriers and high contact resistance. In contrast, strongly metallized vertical contact improves injection but induces MIGS and defects. Lateral (or edge) contact, covalently bonded at the 1D edge, offers the potential for efficient Ohmic contact with reduced FLP, but are limited by fabrication complexity. Finally, high-$\kappa$ dielectrics combined with dual-gate (DG) or gate-all-around (GAA) architectures can maximize gate control, though such geometries impose significant simulation and fabrication challenges.

The transfer characteristics [Fig. \ref{Fig3}(b)] represents the most important information to extract key performance indicators of 2D FETs such as ON-state current and $SS$. An in-depth understanding of the transport properties can be obtained via the local device density of states (LDDOS) [Fig. \ref{Fig3}(c)] and the energy or momentum-resolved transmission function plot [Fig. \ref{Fig3}(d)] \cite{guo2022_TeO2}. 
Figures \ref{Fig3}(e) to (g) shows the 2D FET optimization via multiple parameters including gate length ($L_\mathrm{G}$), source/drain doping concentration, and UL. The $L_\mathrm{G}$ is reduced to meet the density and performance targets. However, aggressive scaling of $L_\mathrm{G}$ intensifies short-channel effects, leading to increased OFF-state leakage current [Fig.~\ref{Fig3}(e)]. Due to the ballistic transport nature of sub-10-nm FET, source/drain doping determines the $n$-type or $p$-type nature and higher doping concentration typically drives higher conduction current in the device [Fig. \ref{Fig3}(f)]. UL physically separates the gate and from the source/drain region. UL weakens the drain-induced barrier lowering (DIBL) effect \cite{khanna2016_DIBL} and SCE, thereby enhancing electrostatic control with steeper $SS$ [Fig. \ref{Fig3}(g)]. However, an overly long UL can reduce the drive current, thus degrading the switching speed of FET. Therefore, UL needs to be carefully optimized for achieving a balance between switching speed and electrostatic control \cite{Huang2021_transistor}. Finally, higher source/drain doping concentration generally increases the drive current through the device, but overly high doping concentration could be challenging to be realistically implemented in 2D semiconductors and could lead to defects and severe scattering effects.

We recommend interested readers to consult a recent review \cite{Quhe2021} on the details of NEGF simulation methods for FET. In the following, we shall focus on reviewing the key performance indicators commonly used to assess the performance of sub-10-nm FET.

\subsection{\label{subsec: MA2Z4_FET} Key Performance Indicators of sub-10-nm Transistor}

\subsubsection{Equivalent Oxide Thickness (EOT)} 

The equivalent oxide thickness (EOT) determines the thickness of the dielectric used in order to achieve the same channel capacitance ($C_\mathrm{Ch}$) as a specific thickness of SiO$_2$. The relationship between EOT and $C_\mathrm{Ch}$ is expressed as 
\begin{equation}
    \mathrm{EOT} = \frac{\varepsilon_{SiO_{2}}}{\varepsilon_{oxide}}t_{ox}
\end{equation}
where $\varepsilon_{SiO_{2}}$ is the permittivity of the SiO$_2$, $\varepsilon_{ox}$ is the permittivity of the dielectric, and $t_{ox}$ is the thickness of the dielectric. Low EOT often benefits from using high-$\kappa$ dielectrics (e.g. HfO$_2$) to maintain high enough $C_\mathrm{Ch}$, while avoiding the excessive leakage currents that occur when using ultrathin SiO$_2$.

\subsubsection{ON-and OFF-state Currents}

The OFF-state current ($I_\mathrm{{OFF}}$) is the leakage current that flows through the FET when it is supposed to be switched off. $I_\mathrm{OFF}$ can originate from: (i) source-to-drain conduction mediated by the Boltzmann tail of the carrier distribution function; and (ii) residue current injected across the gate dielectric between the electrode contact and the channel. Ideally, $I_\mathrm{OFF}$ should be as low as possible so as to minimize unwanted static power dissipation. Suppressing $I_\mathrm{{OFF}}$ is particularly important for low-power (LP) applications where energy efficiency is the core target, but is also tremendously challenging in sub-10-nm devices where the drain-induced barrier lowering (DIBL) effects become more pronounced.

The ON-state current ($I_\mathrm{{ON}}$) is another key figure-of-merit that reflects the transistor’s performance in the active (ON) state.
The $I_\mathrm{{ON}}$ determines a transistor’s ability to conduct charge when turned on by a gate voltage ($V_\mathrm{G}$), thus directly impacting switching speed and circuit performance. Higher $I_\mathrm{{ON}}$ allows faster charging and discharging of circuit nodes, thus enabling higher clock frequencies. Higher $I_\mathrm{{ON}}$ also improves the transistor’s ability to drive capacitive loads in logic circuits.
While $I_\mathrm{{OFF}}$ is typically pre-determined by ITRS or IRDS standards, the $I_\mathrm{{ON}}$ can be determined from the transfer characteristics [Fig. \ref{Fig3}(b)] as the current driven by turn-on voltage ($V_\mathrm{{G,ON}}$),
\begin{equation}
    V_\mathrm{{G,ON}} = V_\mathrm{{G,OFF}} \pm {V_\mathrm{ds}}
\end{equation}
where `+' and `-' is used for \emph{n}-type and \emph{p}-type FETs, respectively, $V_\mathrm{G,OFF}$ is the OFF-state voltage at which the conduction current is driven to the benchmark value of $I_\mathrm{{OFF}}$, and $V_\mathrm{ds}$ is the bias voltage pre-determined by the IRDS and ITRS standards.

A closely related quantity is the current ON/OFF ratio, or $I_\mathrm{{ON}}$/$I_\mathrm{{OFF}}$, which is a key figure of merit for sub-10-nm transistors, indicating the device capability in switching effectively between operating (ON) and non-operating (OFF) states. A high ON/OFF ratio ensures low leakage power, strong signal integrity, and reliable logic operation, which are critical as devices scale down.
Depending on the standards (ITRS or IRDS) and the technology nodes, the ON/OFF ratio typically lies between between $10^3$ and $10^7$ depending on LP or high-performance (HP) applications \cite{ITRS,IRDS_2020}.

\subsubsection{Subthreshold Swing ($SS$)}

The subthreshold swing ($SS$) is a key parameter that measures how efficiently a field-effect transistor (FET) switches from its OFF-state to its ON-state. It is defined as the change in gate voltage ($V_\mathrm{G}$) required to induce a one order of magnitude (i.e., a decade) change in the drain current ($I_\mathrm{d}$) in the subthreshold region:
\begin{equation}
   SS = \frac{ \mathrm{d} V_\mathrm{G}}{ \mathrm{d} \left(\log_{10} I_\mathrm{d}\right)}
\end{equation}
$SS$ is conventionally expressed in units of millivolts per decade (mV~dec$^{-1}$). A \textit{lower} $SS$ value indicates \textit{better gate control} over the channel, enabling the transistor to switch more abruptly between ON and OFF states with a smaller change in gate voltage. This sharper switching characteristic is crucial for reducing power consumption, particularly static power and dynamic power at lower supply voltages.

For conventional FETs operating based on thermionic emission (i.e. the drift-diffusion of carriers over an energy barrier), the $SS$ at room temperature ($T \approx 300$ K) is fundamentally limited by the thermal voltage ($k_B T / q$, where $k_B$ is the Boltzmann constant, $T$ is the absolute temperature, and $q$ is the elementary charge). This limit is given by:
\begin{equation}
    SS_\mathrm{ideal} = \frac{k_B T}{q} \ln(10) \approx 60~\text{mV dec}^{-1}
\end{equation}
This limitation is often referred to as the \textit{Boltzmann tyranny} \cite{salahuddin2008, alam2019_negativecapacitance_review}. In practice, factors such as interface traps and channel doping profiles can degrade $SS$ to values higher than this ideal limit. Furthermore, in highly scaled devices, short-channel effect can severely degrade $SS$, posing a significant challenge for continued transistor miniaturization.

However, the $60~\text{mV dec}^{-1}$ limit is not insurmountable for all transistor types. Devices that operate based on mechanisms other than pure thermionic emission, such as tunnelling FETs (TFETs) \cite{kanungo2022_2DTFET} which utilize quantum mechanical band-to-band tunneling, can potentially achieve a \textit{sub-thermionic subthreshold swing}, i.e. $SS < 60~\text{mV dec}^{-1}$ at room temperature. Such `steep-slope' device is a major research focus for ultralow-power electronics.

The $SS$ is typically extracted from the $IV$-transfer characteristic, specifically from the subthreshold region where the drain current is an exponential function of the gate voltage [Fig. \ref{Fig3}(b)]. It is calculated as the inverse of the maximum slope of the $\log_{10}(I_\mathrm{d})$ versus $V_\mathrm{G}$ curve in this region:
\begin{equation}
    S_\mathrm{min} = \left( \max \left| \frac{ \mathrm{d} \left(\log_{10} I_\mathrm{d}\right)}{ \mathrm{d} V_\mathrm{G}} \right| \right)^{-1}
\end{equation}
Alternatively, an average-valued $SS_\mathrm{avg}$ can be calculated over a specific range of $I_\mathrm{d}$ or $V_\mathrm{G}$ within the subthreshold regime, though the minimum point $SS$ is generally preferred for characterizing the optimal switching sharpness.

\subsubsection{Delay Time}

The delay time ($\tau$) measures how quickly a transistor can switch between ON- and OFF-state, making it a crucial indicator of operating speed. It is given by 
\begin{equation}
    \tau = \frac{C_\mathrm{G} V_\mathrm{ds}}{I_\mathrm{ON}}
\end{equation}
where $C_\mathrm{G}$ is the total gate capacitance calculated as the sum of the fringing capacitance ($C_\mathrm{f}$) and $C_\mathrm{Ch}$, roughly taken to be three times of $C_\mathrm{Ch}$. A lower $\tau$ indicates a faster switching speed, making it desirable for high-frequency and low-latency applications. Optimizing $C_\mathrm{G}$ is essential for achieving high-speed computing, so that $C_\mathrm{G}$ is high enough to boost $I_\mathrm{ON}$, but not so high that it excessively slows down the charging and discharging processes

\subsubsection{Power Delay Product (PDP)}

The power delay product (PDP) quantifies the total energy consumed per switching operation, which can be expressed in two equivalent forms
\begin{subequations}
    \begin{equation}
        \mathrm{PDP} = C_\mathrm{G} V_\mathrm{ds}^{2},
    \end{equation}
    \begin{equation}
        \mathrm{PDP} = \tau I_\mathrm{ON}V_\mathrm{ds}.
    \end{equation}
\end{subequations}
A lower PDP directly translates into lower energy consumption and reduced waste heat generation. Lowering PDP is thus essential for realizing the benefits of increased transistor integration density without prohibitive power consumption. A quantity closely related PDP is the energy-delay product (EDP), 
\begin{equation}
    \mathrm{EDP} = \mathrm{PDP} \times \tau
\end{equation}
EDP simultaneously takes into account both the power consumption and switching speed. While PDP focuses on energy per operation, EDP assesses both energy and performance by penalizing slower devices more heavily through an additional factor of $\tau$, thus providing a more comprehensive assessment on both the energy efficiency and performance of FETs. 

\subsubsection{Commonly used performance metrics requirements of ITRS 2013} 

The ITRS 2013 is the last ITRS roadmap that still clearly outlines the transistor scaling requirement. Due to its clarity in specifying the physical gate length $L_\mathrm{G}$ at each technology nodes, ITRS remains widely used for sub-10-nm FET design. Here we quote the performance metrics target values based on the ITRS 2013 standards. Two types of device requirements are outlined in ITRS 2013, namely for HP and LP applications. For HP applications, the value of $I_\mathrm{OFF}$, $I_\mathrm{ON}$, $I_\mathrm{ON}/I_\mathrm{OFF}$, $\tau$ and PDP are set to 0.1 $\mu$A $\mu$m$^{-1}$, 900 $\mu$A $\mu$m$^{-1}$, $9 \times 10^{3}$, 0.423 ps and 0.24 fF $\mu$m$^{-1}$; whereas for LP applications, the values are set to $5 \times 10^{-5}$ $\mu$A $\mu$m$^{-1}$, 295 $\mu$A $\mu$m$^{-1}$, $5.9 \times 10^{6}$, 1.493 ps and 0.28 fF $\mu$m$^{-1}$, respectively. The EOT for sub-5 nm $L_\mathrm{G}$ FET is usually set to 0.41 nm while the $V_\mathrm{ds}$ is set to 0.64 V.

\subsection{\label{subsec: MA2Z4_FET} A survey of MoSi$_2$N$_4$ family FETs and their $L_\textrm{G}$ scaling limits}

\begin{figure*}
\centering
\includegraphics[width=0.85\textwidth]{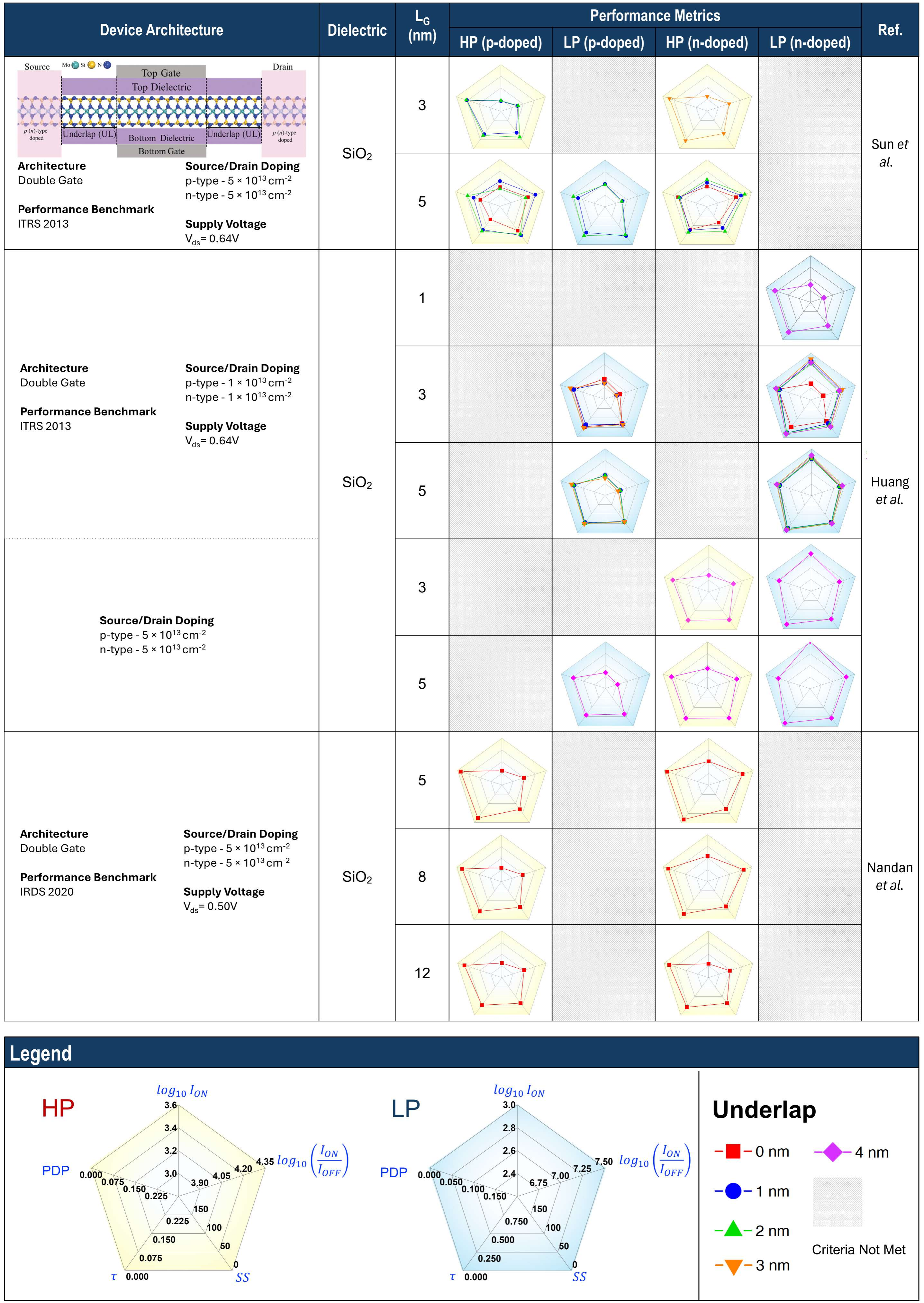}
\caption{\label{Fig4}\textbf{Performance metrics of MoSi$_2$N$_4$ for FET.} The device architecture is either double gate (DG) or single gate (SG), with \emph{n (p)}-type doping at the source and drain simulated using atomic compensation charges. The dielectric material and gate length ($L_\mathrm{G}$) are listed in the columns, while the spider charts display the performance metrics of the devices with different underlap length (UL). For the performance metrics, the logarithm base 10 of $I_\mathrm{ON}$ and $I_\mathrm{ON}/I_\mathrm{OFF}$, subthreshold swing ($SS$), delay time ($\tau$) and power dissipation product (PDP) are displayed using different axes ranges for high-performance (HP) and low-power (LP) applications. Cells shaded in gray denote devices that are unable to meet the ITRS/IRDS requirements. Data taken from - Sun \emph{et al.} \cite{Sun2021_transistor}; Huang \emph{et al.} \cite{Huang2021_transistor}; Nandan \emph{et al.} \cite{Nandan2021_transistor}. } 
\end{figure*}

\begin{figure*}
\centering
\includegraphics[width=0.85\textwidth]{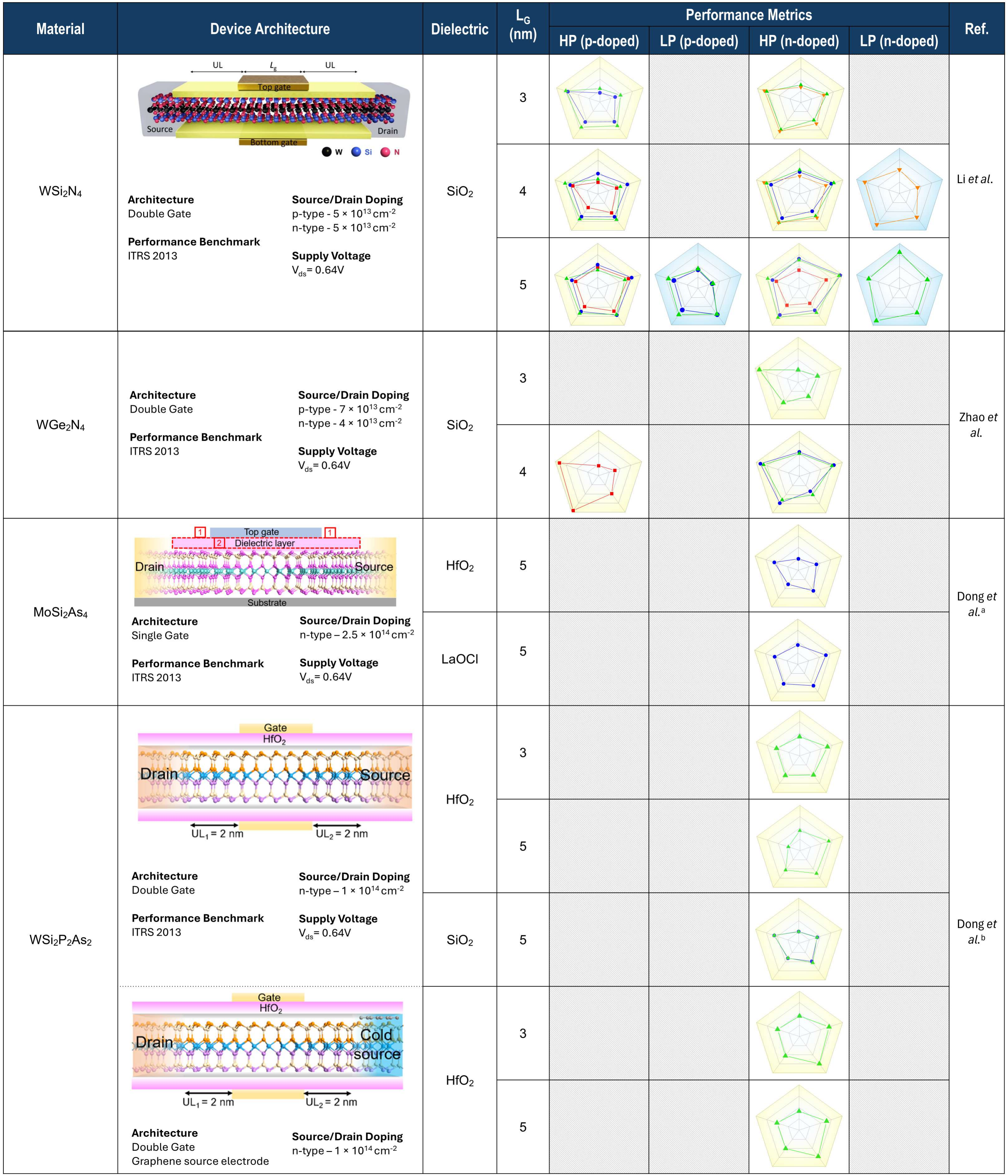}
\caption{\label{Fig5}\textbf{Performance metrics of FETs composed of other members of the MA$_2$Z$_4$ family.} Data taken from - Li \emph{et al.} \cite{li2023_WSi2N4}; Zhao \emph{et al.} \cite{Zhao2021_WGe2N4}; Dong \emph{et al.$^\mathrm{a}$} \cite{Dong2023_MoSi2As4}; Dong \emph{et al.$^\mathrm{b}$} \cite{Dong2023_WSi2P2As2}. } 
\end{figure*}

\begin{figure*}
\centering
\includegraphics[width=0.9\textwidth]{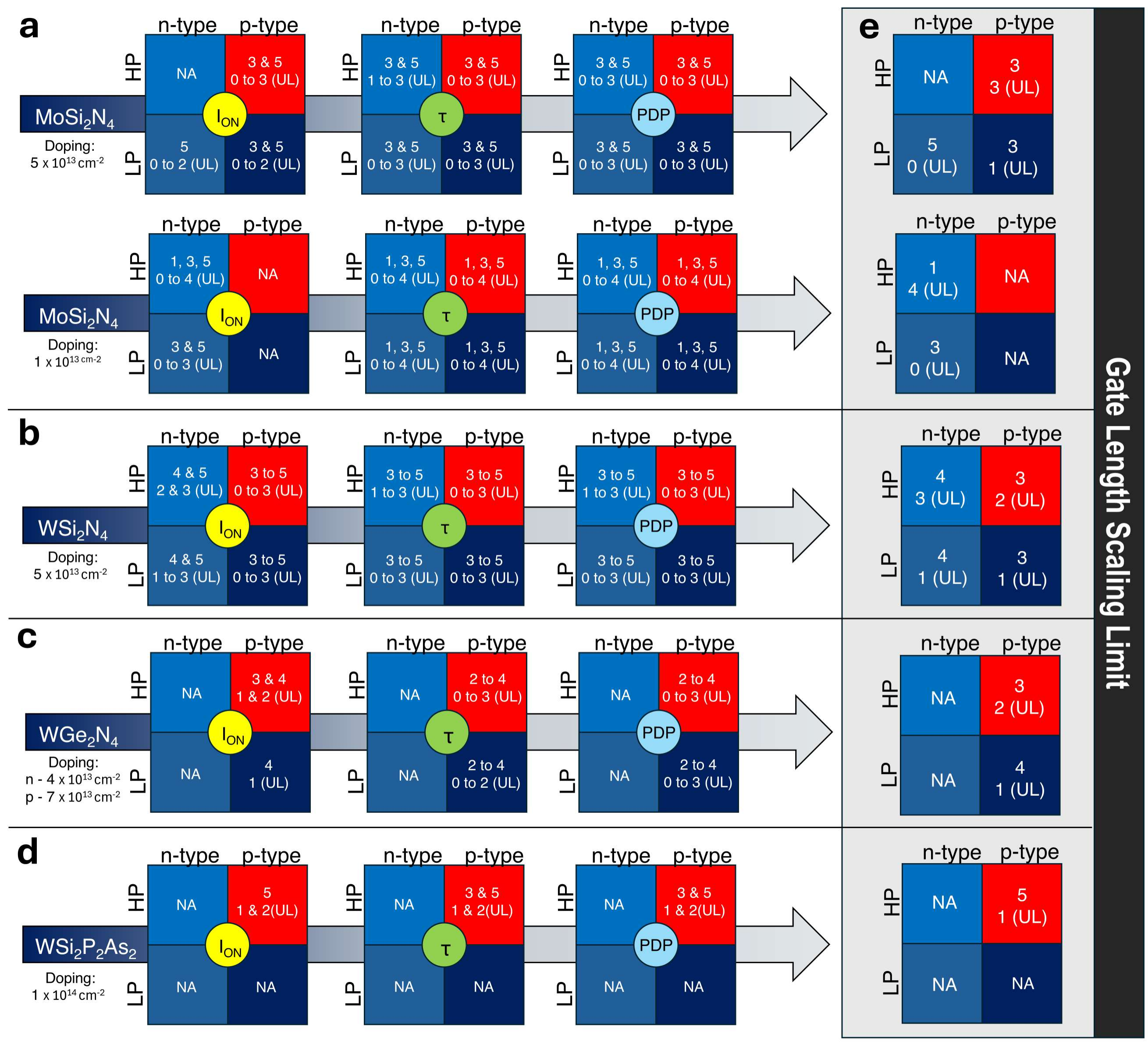}
\caption{\label{Fig6}\textbf{Gate Length scaling limit of MoSi$_2$N$_4$ family FETs.} The smallest $L_\mathrm{G}$ that can still deliver the ITRS 2013 HP/LP application requirements in terms of $I_\mathrm{ON}$, $\tau$, and PDP are listed for (a) MoSi$_2$N$_4$ \cite{Sun2021_transistor, Huang2021_transistor}; (b) WSi$_2$N$_4$ \cite{li2023_WSi2N4}; (c) WGe$_2$N$_4$ \cite{Zhao2021_WGe2N4}; and (d) WSi$_2$P$_2$As$_2$ \cite{Dong2023_WSi2P2As2}. In each cell, the top (bottom) row denotes FET without (with) UL. "NA" denotes instances all reported $L_\mathrm{G}$ fails to meet ITRS requirements. }
\end{figure*}

\begin{figure*}
    \centering
    \includegraphics[width=\linewidth]{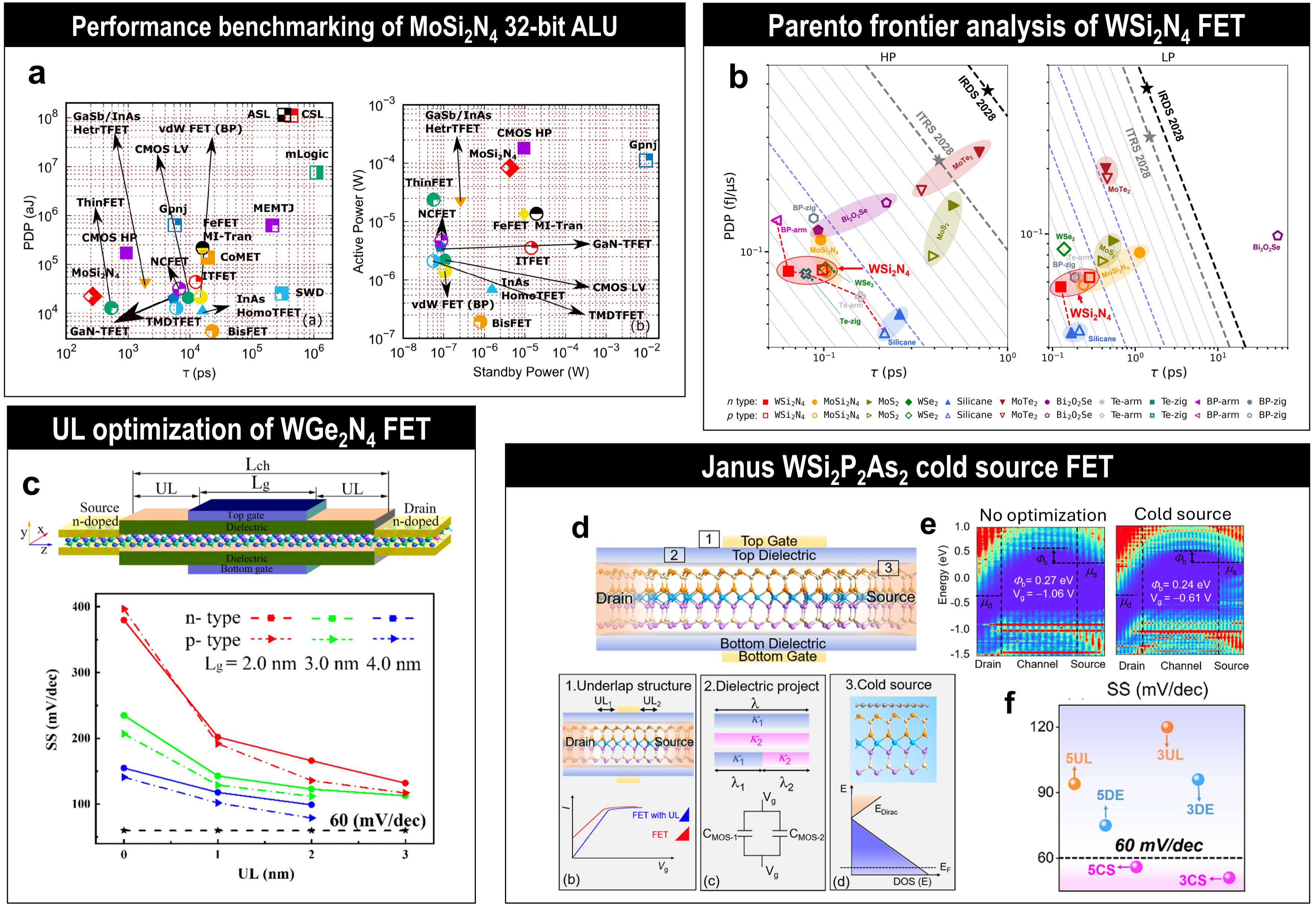}
    \caption{\label{Fig7}\textbf{Performance characteristics of MoSi$_2$N$_4$ family FETs.}(a) The PDP versus $\tau$ and the active versus standby power plots of MoSi$_2$N$_4$-based 32 bit ALU. MoSi$_2$N$_4$ circuit exhibits comparable performance to CMOS counterpart while exhibiting better energy efficiency \cite{nandan2024transistors_MoSi2N4FET}. (b) The $\tau$ versus PDP plot of WSi$_2$N$_4$ and other representative 2D semiconductors shows the strength of WSi$_2$N$_4$ FET in pushing close towards the Pareto frontiers of multiobjective optimization (red dashed line) for both HP and LP applications \cite{li2023_WSi2N4}. (c) Optimization of WGe$_2$N$_4$ FET with UL enables more aggressive $L_\mathrm{G}$ scaling. (d) Cold source, UL and high-$\kappa$ optimization of Janus WSi$_2$P$_2$Aa$_2$ FET. (e) The magnitude of the gate voltage requires to switch OFF the device is significantly reduced when cold source is used, and (f) the $SS$ can be pushed towards the sub-thermionic regime \cite{Dong2023_WSi2P2As2}. (a) Copyright 2021, IEEE; (b) Copyright 2023, American Physical Society; (c) Copyright 2021, American Chemical Society; (d),(e),(f) Copyright 2023, American Chemical Society.}
\end{figure*}

Figures \ref{Fig4} and \ref{Fig5} summarize the performance metrics of best-optimized MA$_2$Z$_4$ family FETs in terms of $I_\textrm{ON}$, $I_\textrm{ON}/I_\textrm{OFF}$, $SS$, $\tau$ and PDP under both HP and LP applications. The detailed numerical values of the performance metrics are listed in TABLE I and II for MoSi$_2$N$_4$, as well as Supplementary TABLE S1, S2 and S3 for other members of the MoSi$_2$N$_4$ family. 
In Fig. \ref{Fig6}, we summarize the gate-length scaling limit, i.e. the \textit{minimum $L_\mathrm{G}$} that can still simultaneously meet the $I_\mathrm{ON}$, $\tau$ and PDP requirements of ITRS 2013, for several representative MoSi$_2$N$_4$ family FETs. Notably, Fig. \ref{Fig6} reveals that WSi$_2$N$_4$ is the only candidate reported thus far that demonstrate device scalability below $L_\textrm{G} = 5$ nm for all configurations of $n$-and $p$-type devices under both HP and LP applications, thereby suggesting its potential as a building block for ultrascaled CMOS technology nodes. Below, we provide an in-depth discussion on the performance of FETs that are based on MoSi$_2$N$_4$ and the other members of MA$_2$Z$_4$.

\subsubsection{MoSi$_2$N$_4$} 

MoSi$_2$N$_4$ FET exhibitis strong potential for HP and LP applications at $L_\mathrm{G}$ = 3, 5 nm \cite{Sun2021_transistor,Huang2021_transistor}. For HP applications,  \emph{n (p)}-type double-gate (DG) MoSi$_2$N$_4$ FETs can satisfy ITRS 2013 requirements down to $L_\mathrm{G}$ = 3 nm, with $I_\mathrm{{ON}}$ in the order of $10^3$ $\mu$A $\mu$m$^{-1}$. This order of current magnitude is comparable to or even higher than theoretical predictions of MoS$_2$ FET \cite{zhao2022_MoS2_Mo_FET, han2014_MoS2_doped_contact_FET}, thus suggesting that MoSi$_2$N$_4$ is a promising 2D channel material in terms of current-carrying capability.     
Source-to-drain tunneling (SDT) is a primary process of leakage current. For MoSi$_2$N$_4$, SDT is a more dominant process in \emph{n}-type FET as compared to \emph{p}-type devices due to the smaller electron effective mass than the hole effecive mass \cite{Nandan2021_transistor}. Hence, the leakage current in \emph{n}-type FET is significantly higher than its \emph{p}-type counterpart. To suppress SDT, DG structures can be used to improve the gate control through an increased $C_\mathrm{G}$ that is twice than that of the SG configuration. 

MoSi$_2$N$_4$ FETs can simultaneously fulfill the $I_\mathrm{ON}$ and $\tau$ requirements under the ITRS 2013 HP and LP applications \cite{Sun2021_transistor, Huang2021_transistor}. This versatile nature of MoSi$_2$N$_4$ FETs outperforms MoS$_2$ FETs which fails to simultaneously fulfill the ITRS HP requirement on $I_\mathrm{ON}$ and $\tau$ \cite{Zhang2021_MoS2}. The faster switching speeds of MoSi$_2$N$_4$ FETs arises from the significantly higher $I_\mathrm{{ON}}$ than MoS$_2$ FETs. MoSi$_2$N$_4$ FETs exhibit a superior EDP = 1.24 $\times$10$^{-30}$ J$\cdot$s $\mu$m$^{-1}$ which outperforms the EDP of MoS$_2$, WSe$_2$, and ReS$_2$ FETs \cite{Sun2021_transistor}, thus suggesting the superior energy efficiency of MoSi$_2$N$_4$ FETs during ON/OFF switching operation. The minimum $SS$ of devices with $L_\mathrm{G}$ = 3, 5 nm can go below the thermionic limit of 60 mV dec$^{-1}$, as the current is dominated by quantum tunneling process instead of thermal injection at the OFF-state. A performance benchmarking of MoSi$_2$N$_4$, in terms of $\tau$ and PDP, with other 2D semicondcutors FET including MoS$_2$, WS$_2$, WSe$_2$ and InSe are shown in Fig. 8. MoSi$_2$N$_4$ FET generally outperforms TMDC FETs, i.e. closer to the bottom-left quadrant of the PDP vs. $\tau$ plot.

It should be noted that using different standards to benchmark FET can lead to drastically different scaling predictions for MoSi$_2$N$_4$. For example, using IRDS 2020 standard \cite{Nandan2021_transistor} of 0.6 nm EOT with $V_\mathrm{ds} = 0.50$ V and $L_\mathrm{G}$ in the range of 3-12 nm, the optimal device under DG \emph{p}-FET configuration can be scaled down to L$_G$ = 5 nm. In contrast, when benchmarked against ITRS 2013, more aggresive scaling limit of $L_\mathrm{G} = 3$ nm is predicted \cite{Sun2021_transistor,Huang2021_transistor}, thus suggesting the importance of clearly outlining the benchmarking standard employed for computational device design.

Beyond the standard FET configuration, negative-capacitance (NC) FET \cite{salahuddin2008} based on MoSi$_2$N$_4$ has also been proposed \cite{Sun2021_transistor}. In NC FET, ferroelectric material is inserted into the gate dielectric to introduce an intrinsic electrical potential that effectively amplifies the surface potential, causing the internal gate voltage to be greater than the applied $V_\mathrm{G}$. This voltage amplification, enabled by the negative capacitance effect, allows the device to overcome the fundamental thermionic limit of 60 mV dec$^{-1}$ for $SS$, thereby achieving sub-thermionic $SS$. For instance, using Hf$_{0.5}$Zr$_{0.5}$O$_{2}$ as a ferroelectric insertion material between the gate and the dielectric has been shown to drastically improve the $I_\text{ON}$ and $SS$ of \emph{n}-type FETs \cite{Sun2021_transistor}. The isolated conduction or valence bands of MoSi$_2$N$_4$ can also be used to realise steep-slope FET with sub-thermionic $SS$ \cite{qu2024_FET_ISB}. A DG MoSi$_2$N$_4$ FET with \emph{p}-type doping concentration of 1 $\times$ 10$^{14}$ cm$^{-2}$ at the source/drain, possess such a characteristic. The ON-state current reaches $I_\mathrm{ON}$ = 1420 and 1005 $\mu$A $\mu$m$^{-1}$ for HP and LP application, respectively, with sub-thermionic $SS < 40$ mV dec$^{-1}$.  

The prowess of transistors lies on their compact integration to form logic gates integrated circuits \cite{wang2012integrated}. Using NEGF device simulations and BSIM4 modelling \cite{khorram2024_MoSi2N4FET}, logic gates composed of $n$-and $p$-type MoSi$_2$N$_4$ FETs, such as NOT, NAND, NOR, XOR, and XNOR, have been demonstrated. Notably, MoSi$_2$N$_4$-based 32-bit adder and arithmetic logic unit (ALU) exhibits similar switching speeds to circuits composed of CMOS HP transistor and 2D-semiconductor-based ThinFET (i.e. WTe$_2$/SnSe$_2$ vertical tunneling FET), while requiring less energy \cite{Nandan2021_transistor} [Fig. \ref{Fig7}(a)]. Their dynamic and standby power consumption is also close to that of CMOS HP, thus positioning MoSi$_2$N$_4$ FETs as a viable building blocks for integrated circuits.

\begin{figure*}
\centering
\includegraphics[width=\textwidth]{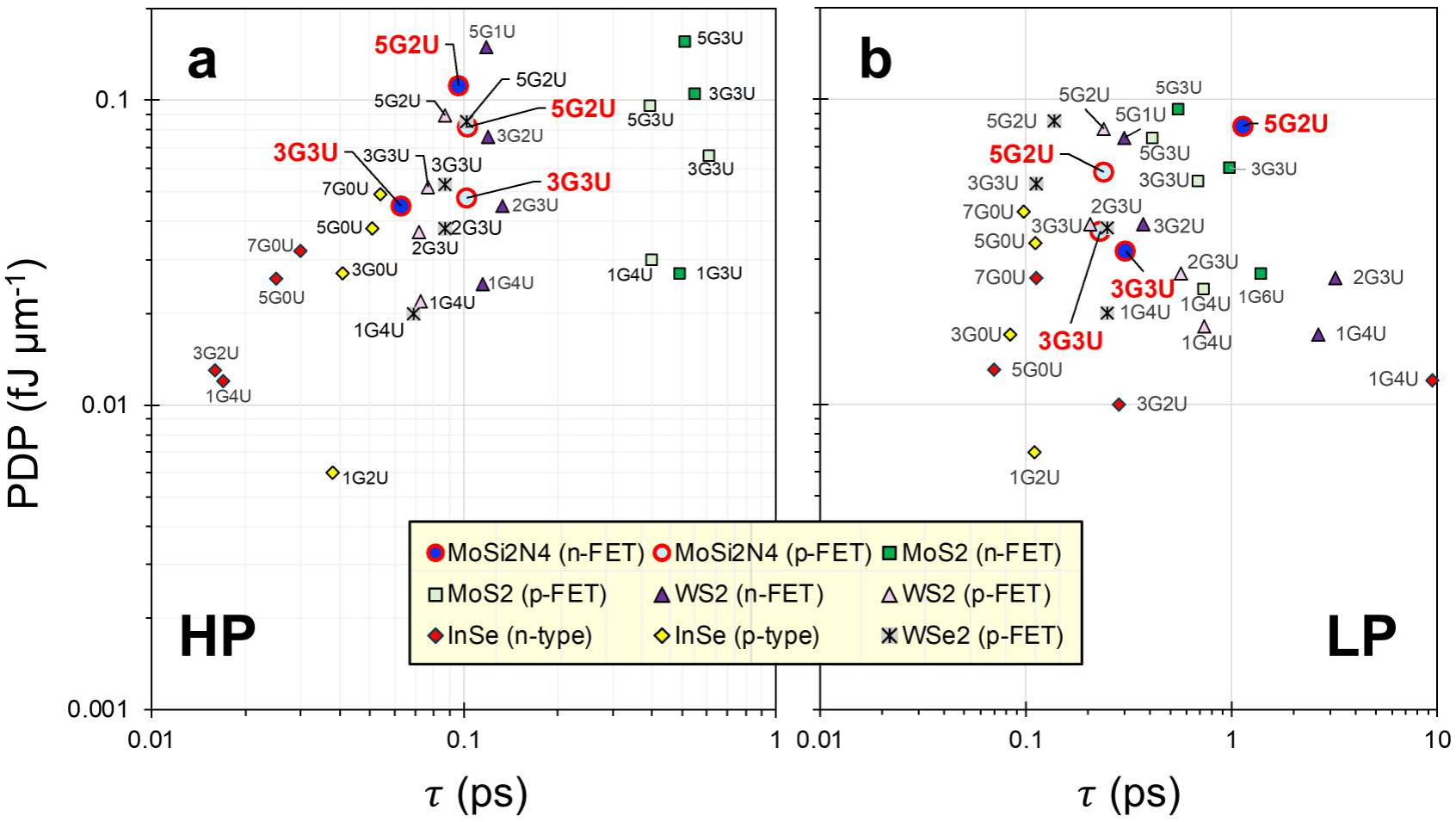}
\caption{\label{Fig8}\textbf{PDP versus delay time ($\tau$) for MoSi$_2$N$_4$ \cite{Sun2021_transistor} bencharmked with MoS$_2$ \cite{Zhang2021_MoS2}, WS$_2$ \cite{sun2025monolayer}, WSe$_2$ \cite{Sun2020_WSe2} and InSe \cite{wang2018many}.} (a) HP and (b) LP device applications. The label $x$G$y$U denotes a transistor with $L_\mathrm{G} = x$ nm and UL = $y$ nm. For each 2D semiconductor, we choose the device with the best EDP for benchmarking. }
\end{figure*}

We benchmark MoSi$_2$N$_4$ \cite{Sun2021_transistor} with representative 2D semiconductors, including MoS$_2$ \cite{Zhang2021_MoS2}, WS$_2$ \cite{sun2025monolayer}, WSe$_2$ \cite{Sun2020_WSe2} and InSe \cite{wang2018many}, in Fig. \ref{Fig8} in terms of PDP and $\tau$ under both HP and LP applications. MoSi$_2$N$_4$ generally outperforms MoS$_2$ and is comparable to WS$_2$. The InSe, which has been demonstrated to exhibit near-ballistic device operation \cite{jiang2023ballistic} due to their excpetional electrical properties \cite{dai2022properties}, remains substantially better than MoSi$_2$N$_4$ in energy efficiencya and switching speed, thus suggesting an open challenge in identifying design strategy or MA$_2$Z$_4$ candidate search that could outperform InSe.

\subsubsection{WSi$_2$N$_4$}

WSi$_2$N$_4$ DG FET [Fig. \ref{Fig3}(e-g)] can be similarly optimized to meet the ITRS 2013 standard for HP and LP applications under suitable source/drain doping concentration, $L_\mathrm{G}$ and UL \cite{li2023_WSi2N4}. For HP applications, DG WSi$_2$N$_4$ FETs satisfy ITRS requirements down to a $L_\mathrm{G}$ = 3 nm, with $I_\mathrm{{ON}}$ values ranging from 1170–2130 $\mu$A $\mu$m$^{-1}$ for \emph{n}-type and 913–1672 $\mu$A/$\mu$m for \emph{p}-type FETs. The ON-state current ratios between the two doping types are found to be in the range of 1.19-1.27, giving a high degree of $np$-symmetry crucial for CMOS applications \cite{li2023_InAs_nanowireFET}. Furthermore, the $SS$ is comparable to that of MoSi$_2$N$_4$, ranging from 69-119 mV dec$^{-1}$ for the \emph{n}-type and 46-59 mV dec$^{-1}$ for \emph{p}-type FETs. The $\tau$ for HP application lies in the range of 0.064–0.112 ps, which is well below the ITRS HP upper limit. The PDP for HP applications are found to be between 0.018–0.084 fJ $\mu$m$^{-1}$, which are significantly lower than the ITRS HP upper limit. 

For LP applications, the scaling limits of \emph{n}-type and \emph{p}-type WSi$_2$N$_4$ FETs are 4 nm and 5 nm, respectively. For LP applications, the \emph{n}-type FETs achieve $I_\mathrm{ON}$ in the range of 417–700 $\mu$A $\mu$m$^{-1}$, significantly surpassing the ITRS lower limit, but the \emph{p}-type FET only achieves a maximum $I_\mathrm{ON}$ of 350 $\mu$A $\mu$m$^{-1}$ which is barely above the ITRS lower limit. The $SS$ for LP applications maintain a value well below the ITRS LP upper limit. The $\tau$ falls within 0.126–0.441 ps, and the PDP stays between 0.016–0.063 fJ $\mu$m$^{-1}$, both satisfying the ITRS criteria. 

The Pareto frontier represents the set of optimal trade-offs in a multi-objective analysis, where improving one metric inevitably degrades another \cite{lotov2008_ParetoFrontier}. In the case of FETs, reducing the delay time $\tau$ increases the switching frequency, which in turn raises the dynamic power consumption, which is the total energy consumed per second during operation \cite{fiori2014_2D_electronics}. To mitigate this, the PDP, which quantifies the energy consumed per switching event, must also be minimized. The Pareto frontier in the $\tau$ vs PDP plot thus marks the boundary of designs that achieve the best compromise between switching speed and energy efficiency. The $L_\mathrm{G} = 5$ nm \emph{n}-type DG WSi$_2$N$_4$ FETs for both HP and LP applications lie on this Pareto frontier, indicating that they deliver one of the most favorable trade-offs among 2D material FETs. Compared to competing materials such as MoSi$_2$N$_4$ \cite{Sun2021_transistor}, MoS$_2$ \cite{Zhang2021_MoS2}, WSe$_2$ \cite{Sun2020_WSe2}, and silicane \cite{pan2020_silicane}, WSi$_2$N$_4$ demonstrates superior overall performance, primarily due to its higher intrinsic carrier mobilities and lower effective masses.

\subsubsection{WGe$_2$N$_4$}

The performance of WGe$_2$N$_4$ DG FETs with $L_\mathrm{G}$ in the range of 2.0-8.8 nm \cite{Zhao2021_WGe2N4} is benchmarked against ITRS 2013 standard. For $L_\mathrm{G}$ above 5 nm devices (5.1 nm, 6.7 nm, 8.8 nm), the performance metrics fulfill the HP lower limit without requiring underlap structure. However, as $L_\mathrm{G}$ shrinks below 5 nm, short-channel effects become significant and severely degrades performance metrics such as $SS$ and $I_\mathrm{{OFF}}$. UL can effectively mitigate the short-channel effect [Fig. \ref{Fig7}(c)]. 
The $L_\mathrm{G}$ can be scaled to 3 (4) nm for \emph{n} (\emph{p})-type FETs. With the use of UL, the switching speed and power consumption of all FETs show a substantial improvement. Specifically, the $\tau$ ranges from 0.11-0.37 (0.18-0.35) ps for \emph{n} (\emph{p})-type devices without underlap, but is reduced to 0.08 (0.14) ps for $L_\mathrm{G}$ = 4 nm and UL = 1 nm. In terms of energy consumption, the PDP of \emph{n} (\emph{p})-type devices shows a downward trend as $L_\mathrm{G}$ scales down to 4 nm, dropping to about 1/6 of the HP upper limit. Further introduction of underlaps for $L_\mathrm{G}$ less than 4 nm reduces the PDP. 

\subsubsection{MoSi$_2$As$_4$}

The impact of four different dielectric materials (SiO$_2$, hBN, HfO$_2$, LaOCl) on the performance of MoSi$_2$As$_4$ SG FETs are evaluated using ITRS 2013 standard \cite{Dong2023_MoSi2As4}.  
At L$_\mathrm{G}$ = 5 nm with SiO$_2$ as the dielectric, the largest $I_\mathrm{ON}$ does not meet the HP lower limit. Using high-$\kappa$ HfO$_2$ dielectric and UL = 1 nm, the $I_\mathrm{{ON}}$ is increased to 1000 $\mu$A $\mu$m$^{-1}$ to meet the HP requirement. 
When using LaOCl as the dielectric, the $I_\mathrm{{ON}}$ further increased to 1380 $\mu$A $\mu$m$^{-1}$. The $SS$ is also significantly improved from 141 mV dec$^{-1}$ (SiO$_2$) to 88 mV dec$^{-1}$ (HfO$_2$) and 83 mV dec$^{-1}$ (LaOCl). The improvement in $I_\mathrm{ON}$ and $SS$ demonstrates that high-$\kappa$ dielectric provides better gate control in MoSi$_2$As$_4$ FET. The use of hBN, which has $\varepsilon_{ox}$ that is close to SiO$_2$, did not significantly improve the device performance, showing only minor changes in $I_\mathrm{ON}$ and $SS$. Overall, HfO$_2$ and LaOCl are identified as promising dielectric for improving the performance of MoSi$_2$As$_4$ SG FETs.

\subsubsection{WSi$_2$P$_2$As$_2$}
For WSi$_2$P$_2$As$_2$ -- a MoSi$_2$N$_4$ family member with Janus morphology \cite{lu2017janus}, various optimization strategies such as UL, high-$\kappa$ dielectric engineering, and cold source implementation \cite{liu2020switching}, are employed to meet the ITRS HP standards under a DG gate configuration \cite{Dong2023_WSi2P2As2} [Fig. \ref{Fig7}(d)]. Using SiO$_2$ as the dielectric, $L_\mathrm{G}$ can only be scaled to 5 nm with $I_\mathrm{{ON}}$ of 1020 $\mu$A $\mu$m$^{-1}$ with UL = 1 nm. 
In contrast, when using HfO$_2$ as the dielectric and UL = 2 nm, L$_G$ can be scaled further to 3 nm with a significantly higher $I_\mathrm{{ON}}$ of 1369 $\mu$A $\mu$m$^{-1}$. The $SS$ is also improved in the case of using HfO$_2$, reaching as low as 75 mV dec$^{-1}$ when L$_G$ = 5 nm and UL = 2 nm. 
Using high-k HfO$_2$ dielectric, graphene cold source electrode to further improve the performance of the device. Unlike conventional metal contacts, which have a high thermal electron distribution, graphene exhibits vanishingly small density of states near the Dirac point \cite{geim2007_rise_graphene, qiu2018_dirac-source_FET}, thus reducing the amount of hot electrons that are injected into the channel during OFF-state. The magnitude of gate voltage needed to switch OFF the device is significantly reduced when both high-$\kappa$ and cold source are incorporated in WSi$_2$P$_2$As$_2$ FET [Fig. \ref{Fig7}(e)]. Importantly, $SS$ can reach the sub-thermionic reimge of 56 mV dec$^{-1}$ at L$_G$ = 5 nm, and 51 mV dec$^{-1}$ at L$_G$ = 3 nm [Fig. \ref{Fig7}(f)], thus suggesting the critical role of cold-source electrode in breaking the Boltzmann tyranny for Janus-WSi$_2$P$_2$As$_2$ FET.

\subsubsection{MA$_2$N$_4$ (M = Ti, Zr, Hf; A = Si, Ge, Sn)} 

The FET performance of the subfamily, MA$_2$N$_4$ (M = Ti, Zr, Hf; A = Si, Ge, Sn), DG FETs have also been explored \cite{hasani2023_FETMA2N4}. The 10 nm $L_\mathrm{G}$ FET exhibits the optimal device performance, with $I_\mathrm{ON}$ in the range of 817.4-1322.6 $\mu$A $\mu$m$^{-1}$, $I_\mathrm{OFF}$ in the range of 165-833 fA and $I_\mathrm{ON}$/$I_\mathrm{OFF}$ ratio in the order of 10$^{9}$. Monolayer TiSi$_2$N$_4$ and HfSi$_2$N$_4$ show the smallest $I_\mathrm{OFF}$ and smallest $SS$ for sub-10-nm channel lengths compared to the other MA$_2$N$_4$ FETs studied in this work, making them the more promising candidates for device downscaling.

\begin{table*}[]
\caption{\label{Performance_Metrics_MoSi2N4_HP} Table of performance metrics for MoSi$_2$N$_4$ FET based on ITRS/IRDS standard for HP applications. Displayed are the gate length ($L_\mathrm{G}$), underlap length (UL), subthreshold swing ($SS$), ON-state current ($I_\mathrm{ON}$), ON/OFF ratio ($I_\mathrm{ON}$/$I_\mathrm{OFF}$), gate capacitance ($C_\mathrm{G}$), delay time ($\tau$) and power dissipation product (PDP). DG/SG denotes double/single gate. For quantities not reported, `-' symbol is used. } 

\resizebox{\textwidth}{!}{%

\begin{tabular}{>{\centering\arraybackslash}m{1.5cm}>{\centering\arraybackslash}m{1.5cm}>{\centering\arraybackslash}m{1.5cm}>{\centering\arraybackslash}m{3cm}>{\centering\arraybackslash}m{1.5cm}>{\centering\arraybackslash}m{1.5cm}>{\centering\arraybackslash}m{1.5cm}>{\centering\arraybackslash}m{2.5cm}>{\centering\arraybackslash}m{2cm}>{\centering\arraybackslash}m{2.5cm}>{\centering\arraybackslash}m{2cm}>{\centering\arraybackslash}m{2cm}>{\centering\arraybackslash}m{2cm}}

% \hline \hline  
% \textbf{Material [Ref.]} & \textbf{Structure} & \textbf{Doping Type} & \textbf{$L_\mathrm{G}$ (nm)}&  \textbf{UL (nm)} & \textbf{SS (mV dec$^{-1}$)} & \textbf{$I_\mathrm{ON}$ ($\mu$A/$\mu$m)} & $I_\mathrm{ON}/I_\mathrm{OFF}$ & \textbf{$C_\mathrm{G}$ (fF $\mu$m$^{-1}$)} & \textbf{$\tau$ (ps)} & \textbf{PDP (fJ $\mu$m$^{-1}$)} \\ \hline \\

\hline \hline  
\textbf{Benchmark} & \textbf{[Ref.]} & \textbf{Structure} & \textbf{Doping Type} & \textbf{Dielectric} &  $\boldsymbol{L_\mathrm{G}}$ &  $\boldsymbol{UL}$ & $\boldsymbol{SS}$ &  $\boldsymbol{I_\mathrm{ON}}$ &  $\boldsymbol{I_\mathrm{ON}/I_\mathrm{OFF}}$ & $\boldsymbol{C_\mathrm{G}}$ &  $\boldsymbol{\tau}$ & \textbf{PDP} \\ 

& &  & \textbf{\& Concentration} & & (nm) & (nm) & (mV dec$^{-1}$) & ($\mu$A $\mu$m$^{-1}$) & & (fF $\mu$m$^{-1}$) & (ps) & (fJ $\mu$m$^{-1}$) \\

\hline \\

ITRS 2013 & \cite{Sun2021_transistor} & DG & \emph{n}-Type & SiO$_2$ & 1 & 0 & 1044 & - & - & - & - & - \\
& & & (5 $\times 10^{13}$ cm$^{-2}$) & & & 1 & 593 & - & - & - & - & - \\
& & & & & & 2 & 209 & 59  & 5.90 $\times 10^{2}$  & 0.056  & 0.605  & 0.023  \\
& & & & & & 3 & 144 & 318  & 3.18 $\times 10^{3}$  & 0.048  & 0.096  & 0.020   \\
& & & & & & 4 & 118 & 490  & 4.90 $\times 10^{3}$  & 0.048  & 0.062  & 0.019  \\
& & & & & 3 & 0 & 176 & 120  & 1.20 $\times 10^{3}$  & 0.182  & 0.972  & 0.075  \\
& & & & & & 1 & 140 & 470  & 4.70 $\times 10^{3}$  & 0.171  & 0.233  & 0.070 \\
& & & & & & 2 & 97 & 800  & 8.00 $\times 10^{3}$  & 0.152  & 0.122  & 0.062  \\
& & & & & & 3 & 80 & 1112  & 1.11 $\times 10^{4}$  & 0.110  & 0.063  & 0.045  \\
& & & & & 5 & 0 & 115 & 1382  & 1.38 $\times 10^{4}$  & 0.257  & 0.119  & 0.105 \\
& & & & & & 1 & 87 & 1613  & 1.61 $\times 10^{4}$  & 0.286  & 0.113  & 0.117  \\
& & & & & & 2 & 69 & 1813 & 1.81 $\times 10^{4}$  & 0.272  & 0.096  & 0.112  \\ \\

& & & \emph{p}-Type & & 1 & 0 & 377 & - & - & - & - & - \\
& & & (5 $\times 10^{13}$ cm$^{-2}$) & & & 1 & 179  & 114  & 1.14 $\times 10^{3}$  & 0.071  & 0.398  & 0.029  \\
& & & & & & 2 & 116  & 290  & 2.90 $\times 10^{3}$ & 0.053 & 0.117 & 0.022 \\
& & & & & & 3 & 79  & 393  & 3.93 $\times 10^{3}$ & 0.056 & 0.092 & 0.023 \\
& & & & & & 4 & 59  & 362  & 3.62 $\times 10^{3}$ & 0.041 & 0.073 & 0.017 \\
& & & & & 3 & 0 & 113  & 573  & 5.73 $\times 10^{3}$ & 0.262 & 0.292 & 0.107 \\
& & & & & & 1 & 86  & 940  & 9.40 $\times 10^{3}$ & 0.171 & 0.117 & 0.070 \\
& & & & & & 2 & 63  & 959  & 9.59 $\times 10^{3}$ & 0.161 & 0.107 & 0.066 \\
& & & & & & 3 & 47  & 734  & 7.34 $\times 10^{3}$ & 0.117 & 0.102 & 0.048 \\
& & & & & 5 & 0 & 71 & 1343 & 1.34 $\times 10^{4}$ & 0.410 & 0.195 &  0.168 \\
& & & & & & 1 & 46 & 1690 & 1.69 $\times 10^{4}$ & 0.297 & 0.113 & 0.122 \\
& & & & & & 2 & 52 & 1244 & 1.24 $\times 10^{4}$ & 0.201 & 0.103 & 0.082 \\ \\

\hline \\

ITRS 2013 & \cite{Huang2021_transistor} & DG & \emph{n}-Type & SiO$_2$ & 1 & 0 & 166.164 & 423 & 4.23 $\times 10^{3}$ & 0.1194 & 0.181 & 0.049 \\
& & & (1 $\times 10^{13}$ cm$^{-2}$) & & & 1 & 127.24 & 653 & 6.53 $\times 10^{3}$ & 0.1403 & 0.138 & 0.057 \\
& & & & & & 2 & 104.855 & 682  & 6.82 $\times 10^{3}$ & 0.1242  & 0.117  & 0.051  \\
& & & & & & 3 & 97.128 & 777 & 7.77 $\times 10^{3}$ & 0.1087 & 0.090 & 0.045 \\
& & & & & & 4 & 74.818 & 766 & 7.66 $\times 10^{3}$ & 0.0993 & 0.083 & 0.041 \\
& & & & & 3 & 0 & 86.414  & - & - & - & - & - \\
& & & & & & 1 & 74.077 & 671 & 6.71 $\times 10^{3}$ & 0.1419 & 0.135 & 0.058 \\
& & & & & & 2 & 64.429 & 683 & 6.83 $\times 10^{3}$ & 0.1276 & 0.120 & 0.052 \\
& & & & & & 3 & 57.623 & 714 & 7.14 $\times 10^{3}$ & 0.1135 & 0.102 & 0.046 \\
& & & & & & 4 & 57.240 & 810 & 8.10 $\times 10^{3}$ & 0.1058 & 0.084 & 0.043 \\
& & & & & 5 & 0 & 57.006  & 666 & 6.66 $\times 10^{3}$ & 0.1326 & 0.127 & 0.054 \\
& & & & & & 1 & 56.666  & - & - & - & - & - \\
& & & & & & 2 & 53.980  & 673 & 6.73 $\times 10^{3}$ & 0.1259 & 0.120 & 0.052 \\
& & & & & & 3 & 51.641  & 702 & 7.02 $\times 10^{3}$ & 0.1119 & 0.102 & 0.046 \\
& & & & & & 4 & 51.157  & 817 & 8.17 $\times 10^{3}$ & 0.1039 & 0.081 & 0.043 \\ \\

& & & \emph{p}-Type &  & 1 & 0 & 118.811 & - & - & - & - & - \\
& & & (1 $\times 10^{13}$ cm$^{-2}$) & & & 1 & 91.958 & 321 & 3.21 $\times 10^{3}$ & 0.1447 & 0.288 & 0.059 \\
& & & & & & 2 & 84.386 & 330 & 3.30 $\times 10^{3}$ & 0.1331 & 0.258 & 0.055  \\
& & & & & & 3 & 82.238 & 348 & 3.48 $\times 10^{3}$ & 0.1089 & 0.200 & 0.045  \\
& & & & & & 4 & 75.820 & 392 & 3.92 $\times 10^{3}$ & 0.1021 & 0.167 & 0.042  \\
& & & & & 3 & 0 & 70.371 & - & - & - & - & - \\
& & & & & & 1 & 66.086 & 324 & 3.24 $\times 10^{3}$ & 0.1489 & 0.294 & 0.061 \\
& & & & & & 2 & 65.927 & 323 & 3.23 $\times 10^{3}$ & 0.1297 & 0.257 & 0.053 \\
& & & & & & 3 & 64.414 & 336 & 3.36 $\times 10^{3}$ & 0.1129 & 0.215 & 0.046 \\
& & & & & & 4 & 63.689 & 403 & 4.03 $\times 10^{3}$ & 0.1078 & 0.171 & 0.044 \\
& & & & & 5 & 0 & 64.957 & 402 & 4.02 $\times 10^{3}$ & 0.1436 & 0.229 & 0.059 \\
& & & & & & 1 & 60.251 & 324 & 3.24 $\times 10^{3}$ & 0.1467 & 0.290 & 0.060 \\
& & & & & & 2 & 59.168 & 301 & 3.01 $\times 10^{3}$ & 0.1365 & 0.290 & 0.056 \\
& & & & & & 3 & 58.717 & 332 & 3.32 $\times 10^{3}$ & 0.1185 & 0.228 & 0.049 \\
& & & & & & 4 & 57.524 & 391 & 3.91 $\times 10^{3}$ & 0.1073 & 0.176 & 0.044 \\ \\

& & & \emph{n}-Type & & 1 & 4 & 74.244 & 866 & 8.66 $\times 10^{3}$ & 0.1296 & 0.096 & 0.053 \\ 
& & & (5 $\times 10^{13}$ cm$^{-2}$) & & 3 & 4 & 51.758 & 1206 & 1.206 $\times 10^{3}$ & 0.1387 & 0.074 & 0.057 \\ 
& & & & & 5 & 4 & 43.991 & 1390 & 1.39 $\times 10^{3}$ & 0.1384 & 0.064 & 0.057 \\ \\

& & & \emph{p}-Type & & 1 & 4 & 83.520 & 448 & 4.48 $\times 10^{3}$ & 0.1236 & 0.177 & 0.051 \\ 
& & & (5 $\times 10^{13}$ cm$^{-2}$) & & 3 & 4 & 69.381 & 592 & 5.92 $\times 10^{3}$ & 0.1345 & 0.145 & 0.055 \\ 
& & & & & 5 & 4 & 63.901 & 618 & 6.18 $\times 10^{3}$ & 0.135 & 0.140 & 0.055 \\ \\

\hline \\

IRDS 2020 & \cite{Nandan2021_transistor} & DG & \emph{n}-Type & SiO$_2$ & 3 & 0  & 118.6 & 663 & 6.630 $\times 10^{3}$ & 0.028 & 0.022 & 0.007 \\
& & & & & 5 & 0 & 73.5 & 1610 & 1.61 $\times 10^{4}$ & 0.08 & 0.027 & 0.02 \\
& & & & & 8 & 0 & 66.2 & 1720 & 1.72 $\times 10^{4}$ & 0.14 & 0.041 & 0.035 \\
& & & & & 12 & 0 & 65.4 & 1750 & 1.75 $\times 10^{4}$ & 0.204 & 0.058 & 0.051 \\ \\

& & & \emph{p}-Type & & 3 & 0 & 80.0 & 812 & 8.12 $\times 10^{3}$ & 0.052 & 0.031 & 0.013 \\
& & & & & 5 & 0 & 70.5 & 1112 & 1.112 $\times 10^{4}$ & 0.08 & 0.036 & 0.02 \\
& & & & & 8 & 0 & 64.5 & 1091 & 1.091 $\times 10^{4}$ & 0.14 & 0.064 & 0.035 \\
& & & & & 12 & 0 & 63.6 & 1122 & 1.122 $\times 10^{4}$ & 0.184 & 0.08 & 0.046 \\ \\

& & SG & \emph{n}-Type & & 3 & 0 & - & - & - & - & - & - \\
& &  &  & & 5 & 0 & 107 & 581 & 5.81 $\times 10^{3}$ & - & - & - \\
& &  &  & & 8 & 0 & 83 & 800 & 8.00 $\times 10^{3}$ & - & - & - \\
& &  &  & & 12 & 0 & 73.7 & 1072 & 1.072 $\times 10^{4}$ & - & - & - \\ \\

& &  & \emph{p}-Type & & 3 & 0 & 107 & 302 & 3.02 $\times 10^{3}$ & - & - & - \\
& &  &  & & 5 & 0 & 83 & 455 & 4.55 $\times 10^{3}$ & - & - & - \\
& &  &  & & 8 & 0 & 74 & 588 & 5.88 $\times 10^{3}$ & - & - & - \\
& &  &  & & 12 & 0 & 67.6 & 596 & 5.96 $\times 10^{3}$ & - & - & - \\ \\

\hline \hline
\end{tabular}
}
\end{table*}

\begin{table*}[]
% \captionsetup{labelformat=empty} % Removes "TABLE X"
\caption{\label{Performance_Metrics_MoSi2N4_LP} Table of performance metrics for MoSi$_2$N$_4$ FET based on ITRS standard for LP applications.} 

\resizebox{\textwidth}{!}{%

\begin{tabular}{>{\centering\arraybackslash}m{1.5cm}>{\centering\arraybackslash}m{1.5cm}>{\centering\arraybackslash}m{1.5cm}>{\centering\arraybackslash}m{3cm}>{\centering\arraybackslash}m{1.5cm}>{\centering\arraybackslash}m{1.5cm}>{\centering\arraybackslash}m{1.5cm}>{\centering\arraybackslash}m{2.5cm}>{\centering\arraybackslash}m{2cm}>{\centering\arraybackslash}m{2.5cm}>{\centering\arraybackslash}m{2cm}>{\centering\arraybackslash}m{2cm}>{\centering\arraybackslash}m{3cm}}

\hline \hline  
\textbf{Benchmark} & \textbf{[Ref.]} & \textbf{Structure} & \textbf{Doping Type} & \textbf{Dielectric} &  $\boldsymbol{L_\mathrm{G}}$ &  $\boldsymbol{UL}$ & $\boldsymbol{SS}$ &  $\boldsymbol{I_\mathrm{ON}}$ &  $\boldsymbol{I_\mathrm{ON}/I_\mathrm{OFF}}$ & $\boldsymbol{C_\mathrm{G}}$ &  $\boldsymbol{\tau}$ & \textbf{PDP} \\ 

& &  & \textbf{\& Concentration} & & (nm) & (nm) & (mV dec$^{-1}$) & ($\mu$A $\mu$m$^{-1}$) & & (fF $\mu$m$^{-1}$) & (ps) & (fJ $\mu$m$^{-1}$) \\

\hline \\

ITRS 2013 & \cite{Sun2021_transistor} & DG & \emph{n}-Type & SiO$_2$ & 1 & 0 & 1044 & - & - & - & - & - \\
& & & (5 $\times 10^{13}$ cm$^{-2}$) & & & 1 & 593 & - & - & - & - & - \\
& & & & & & 2 & 209 &  - &  - &  - &  - &  -  \\
& & & & & & 3 & 144 &  1 &  2.00 $\times 10^{4}$ & 0.042 & 26.612 & 0.017  \\
& & & & & & 4 & 118 &  18 & 3.60 $\times 10^{5}$ & 0.038 & 1.355 & 0.016 \\
& & & & & 3 & 0 & 176 &  - &  - &  - &  - &  -  \\
& & & & & & 1 & 140 &  - &  - &  - &  - &  - \\
& & & & & & 2 & 97 &  20 &  4.00 $\times 10^{5}$ &  0.100 & 3.189 & 0.041 \\
& & & & & & 3 & 80 &  163 &  3.26 $\times 10^{6}$ &  0.077 &  0.303 &  0.032 \\
& & & & & 5 & 0 & 115 &  2 &  4.00 $\times 10^{4}$ &  0.220 &  70.385 &  0.090 \\
& & & & & & 1 & 87 &  114 &  2.28 $\times 10^{6}$ &  0.201 &  1.130 &  0.082 \\
& & & & & & 2 & 69 &  77 &  1.54 $\times 10^{6}$ &  0.175 &  1.457 &  0.072 \\ \\

& & & \emph{p}-Type & & 1 & 0 & 377 & - & - & - & - & - \\
& & & (5 $\times 10^{13}$ cm$^{-2}$) & & & 1 & 179  & - &  - &  - & - & - \\
& & & & & & 2 & 116 & 6 & 1.20 $\times 10^{5}$ & 0.049 & 5.192 & 0.020 \\
& & & & & & 3 & 79 & 25 & 5.00 $\times 10^{5}$ & 0.041 & 1.060 & 0.017 \\
& & & & & & 4 & 59 & 46 & 9.20 $\times 10^{5}$ & 0.039 & 0.539 & 0.016 \\
& & & & & 3 & 0 & 113 & 4 & 8.00 $\times 10^{5}$ & 0.176 & 28.090 & 0.072 \\
& & & & & & 1 & 86 & 67 & 1.34 $\times 10^{6}$ & 0.134 & 1.278 & 0.055 \\
& & & & & & 2 & 63 & 249 & 4.98 $\times 10^{6}$ & 0.120 & 0.308 & 0.049 \\
& & & & & & 3 & 47 & 252 & 5.04 $\times 10^{6}$ & 0.091 & 0.230 & 0.037 \\
& & & & & 5 & 0 & 71 & 213 & 4.26 $\times 10^{6}$ & 0.305 & 0.916 & 0.125 \\
& & & & & & 1 & 46 & 390 & 7.80 $\times 10^{6}$ & 0.194 & 0.318 & 0.079 \\
& & & & & & 2 & 52 & 378 & 7.56 $\times 10^{6}$ & 0.140 & 0.238 & 0.058 \\ \\

\hline \\

ITRS 2013 & \cite{Huang2021_transistor} & DG & \emph{n}-Type & SiO$_2$ & 1 & 0 & 166.164 & - & - & - & - & - \\
& & & (1 $\times 10^{13}$ cm$^{-2}$) & & & 1 & 127.24 & - & - & - & - & - \\
& & & & & & 2 & 104.885 & 44 & 8.80 $\times 10^{5}$ & 0.1242 & 1.814 & 0.051 \\
& & & & & & 3 & 97.128 & 231 & 4.62 $\times 10^{6}$ & 0.1087 & 0.301 & 0.045 \\
& & & & & & 4 & 74.818 & 315 & 6.30 $\times 10^{6}$ & 0.0993 & 0.202 & 0.041 \\
& & & & & 3 & 0 & 86.414 & 299 & 5.98 $\times 10^{6}$ & 0.131 & 0.280 & 0.054 \\
& & & & & & 1 & 74.077 & 743 & 1.49 $\times 10^{7}$ & 0.1419 & 0.122 & 0.058 \\
& & & & & & 2 & 64.429 & 659 & 1.32 $\times 10^{7}$ & 0.1276 & 0.124 & 0.052 \\
& & & & & & 3 & 57.623 & 793 & 1.59 $\times 10^{7}$ & 0.1135 & 0.092 & 0.046 \\
& & & & & & 4 & 57.240 & 685 & 1.37 $\times 10^{7}$ & 0.1058 & 0.099 & 0.043 \\
& & & & & 5 & 0 & 57.006 & 714 & 1.43 $\times 10^{7}$ & 0.1326 & 0.119 & 0.054 \\
& & & & & & 1 & 56.666 & 684 & 1.37 $\times 10^{7}$ & 0.1371 & 0.128 & 0.056 \\
& & & & & & 2 & 53.980 & 689 & 1.38 $\times 10^{7}$ & 0.1259 & 0.117 & 0.052 \\
& & & & & & 3 & 51.641 & 750 & 1.50 $\times 10^{7}$ & 0.1119 & 0.095 & 0.046 \\
& & & & & & 4 & 51.157 & 787 & 1.57 $\times 10^{7}$ & 0.1039 & 0.084 & 0.043 \\ \\

& & & \emph{p}-Type & & 1 & 0 & 118.811 & - & - & - & - & - \\
& & & (1 $\times 10^{13}$ cm$^{-2}$) & & & 1 & 91.958 & - & - & - & - & - \\
& & & & & & 2 & 84.386 & 202 & 4.40 $\times 10^{6}$ & 0.1331 & 0.422 & 0.055 \\
& & & & & & 3 & 82.238 & 168 & 3.36 $\times 10^{6}$ & 0.1089 & 0.415 & 0.045 \\
& & & & & & 4 & 75.820 & 120 & 2.40 $\times 10^{6}$ & 0.1021 & 0.545 & 0.042 \\
& & & & & 3 & 0 & 70.371 & 355 & 7.10 $\times 10^{6}$ & 0.1469 & 0.265 & 0.060 \\
& & & & & & 1 & 66.086 & 302 & 6.04 $\times 10^{6}$ & 0.1489 & 0.316 & 0.061 \\
& & & & & & 2 & 65.927 & 278 & 5.56 $\times 10^{6}$ & 0.1297 & 0.300 & 0.053 \\
& & & & & & 3 & 64.414 & 300 & 6.00 $\times 10^{6}$ & 0.1129 & 0.241 & 0.046 \\
& & & & & & 4 & 63.689 & 216 & 4.32 $\times 10^{6}$ & 0.1078 & 0.319 & 0.044 \\
& & & & & 5 & 0 & 64.957 & 227 & 4.54 $\times 10^{6}$ & 0.1436 & 0.405 & 0.059 \\
& & & & & & 1 & 60.251 & 348 & 6.96 $\times 10^{6}$ & 0.1467 & 0.270 & 0.060 \\
& & & & & & 2 & 59.168 & 342 & 6.84 $\times 10^{6}$ & 0.1365 & 0.255 & 0.056 \\
& & & & & & 3 & 58.717 & 312 & 6.24 $\times 10^{6}$ & 0.1185 & 0.243 & 0.049 \\
& & & & & & 3 & 57.524 & 233 & 4.66 $\times 10^{6}$ & 0.1073 & 0.295 & 0.044 \\ \\

& & & \emph{n}-Type & & 1 & 4 & 74.244 & - & - & - & - & - \\ 
& & & (5 $\times 10^{13}$ cm$^{-2}$) & & 3 & 4 & 51.758 & 691 & 1.382 $\times 10^{7}$ & 0.1387 & 0.120 & 0.057 \\ 
& & & & & 5 & 4 & 43.991 & 1025 & 2.05 $\times 10^{7}$ & 0.1384 & 0.086 & 0.057 \\ \\

& & & \emph{p}-Type & & 1 & 4 & 83.520 & - & - & - & - & - \\ 
& & & (5 $\times 10^{13}$ cm$^{-2}$) & & 3 & 4 & 69.381 & 266 & 5.32 $\times 10^{6}$ & 0.1345 & 0.324 & 0.055 \\ 
& & & & & 5 & 4 & 63.901 & 295 & 5.90 $\times 10^{6}$ & 0.135 & 0.293 & 0.055 \\ \\

\hline \hline
\end{tabular}
}
\end{table*}

\begin{figure*}
\centering
\includegraphics[width=\textwidth]{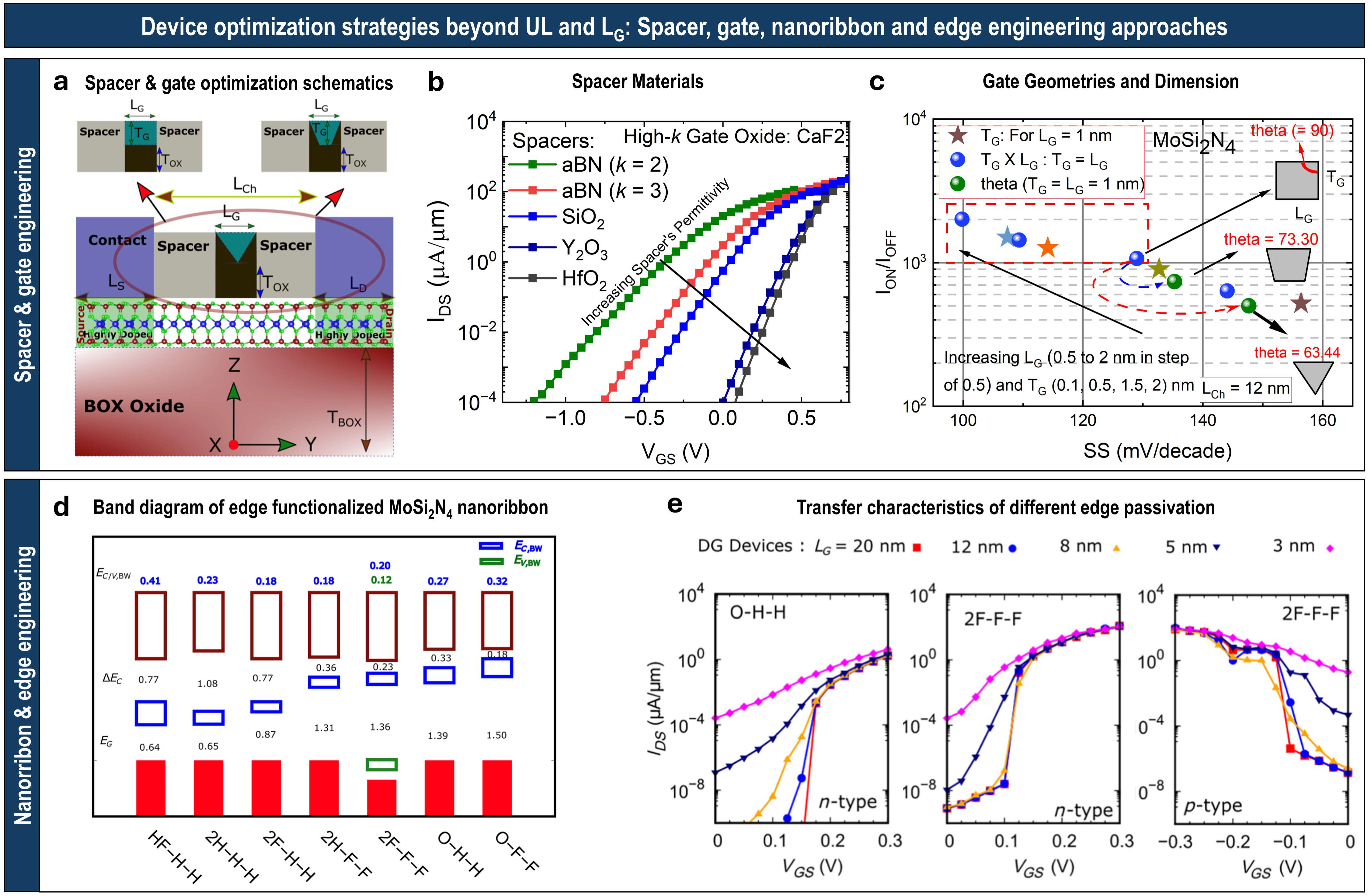}
\caption{\label{Fig9}\textbf{Optimization strategies beyond UL and $L_\mathrm{G}$.} (a) Schematic of MoSi$_2$N$_4$ FET that uses spacer, BOX oxide and gate with different structures \cite{nandan2024transistors_MoSi2N4FET}. (b) transfer characteristic of MoSi$_2$N$_4$ FET, with CaF$_2$ as the gate dielectric and several other materials as the spacer \cite{nandan2024transistors_MoSi2N4FET}. (c) Role of gate thickness ($T_\mathrm{G}$, $L_\mathrm{G}$ and geometry on the performance of MoSi$_2$N$_4$ FET \cite{nandan2024transistors_MoSi2N4FET}. (d) Electronic band diagrams of possible combinations of edge functionalized MoSi$_2$N$_4$ nanoribbon. Blue (Green) boxes denotes the conduction (valence) bandwidth, respectively \cite{nandan2023_EtchPassivationFET}. (e) transfer characteristic of edge-passivated MoSi$_2$N$_4$ nanoribbon FET of various $L_\mathrm{G}$ \cite{nandan2023_EtchPassivationFET}. (a),(b),(c) Copyright 2025, IEEE; (d),(e) Copyright 2023, American Physical Society.}
\end{figure*}

\subsection{Performance Optimization of MoSi$_2$N$_4$ FET}

Besides device optimization via $L_\mathrm{G}$ and UL, other approaches such as spacer materials \cite{nandan2024transistors_MoSi2N4FET}, gate geometry \cite{nandan2024transistors_MoSi2N4FET}, edge passivation \cite{nandan2023_EtchPassivationFET}, and strain modulation \cite{ghobadi2022_StrainedFET} have also been employed to engineer the carrier transport, subthreshold behavior, and overall energy efficiency of MA$_2$Z$_4$ FETs. Focusing on MoSi$_2$N$_4$ FETs, we review how such strategies can be used to further improve the device performance.

\subsubsection{Spacer and gate engineering} 

The influence of spacer material and gate geometry for $L_\mathrm{G}$ = 1 nm MoSi$_2$N$_4$ FET has been computationally investigated \cite{nandan2024transistors_MoSi2N4FET} using a combination of DFT, maximally localised Wannier functions (MLWFs) and NEGF [Fig. \ref{Fig9}(a)]. The device employs a SG configuration as implemented via bottom SiO$_2$ gate dielectric. Different spacer materials including air, SiO$_2$, aBN, Y$_2$O$_3$, HfO$_2$ , CaF$_2$) and dielectric (SiO$_2$, CaF$_2$) are also included. When using spacer materials with increasing dielectric permittivity, the gate electrostatic over the channel is enhanced with lowered $SS$ [Fig. \ref{Fig9}(b)]. This improved gate electrostatic is a result of an increased fringing capacitance ($C_\mathrm{f}$) arising from the fringing field from the gate, given by 
\begin{equation}
   C_\mathrm{f}=\epsilon_\mathrm{spacer} \mathrm{ln}\left(1+\frac{T_\mathrm{G}}{T_\mathrm{ox}}\right),  
\end{equation}
where $\epsilon_\mathrm{spacer}$ is the electric permittivity of spacer, $T_\mathrm{G}$ and $T_\mathrm{ox}$ is the thickness of the gate and dielectric, respectively. Increasing $T_\mathrm{ox}$ in the range of 1.1-3.3 nm while using the same dielectric, clearly shows a decrease in $C_\mathrm{f}$, thereby deteriorating the gate's control. The influence of gate metal dimension and geometry on FET performance has also been examined [Fig. \ref{Fig9}(c)]. Uniformly increasing the gate area lowers the $SS$, whereas altering the shape of the metal gate from square to triangular degrades the device performance, thus suggesting that simple gate geometry is sufficient for device design.

\subsubsection{Nanoribbon and edge engineering} 

The zigzag nanoribbon MoSi$_2$N$_4$ FETs with SG and DG configurations have been investigated with different edge passivation of H, F, N and O atoms \cite{nandan2023_EtchPassivationFET}. The edge-passivated MoSi$_2$N$_4$ exhibit narrow band-width conduction band and the nanoribbon with purely F atoms passivation further exhibits narrow band-width valence band [Fig. \ref{Fig9} (d)], making them suitable for energy-efficient sub-thermionic operations. From the transfer characteristics [Fig. \ref{Fig9}(e)], the average $SS$ of some edge-passivated devices can reach values much smaller than the sub-thermionic limit of 60 mV dec$^{-1}$ (e.g. $SS_\mathrm{ave}$ $<$ 20 mV dec$^{-1}$ in \emph{n}-type fully F atoms terminated MoSi$_2$N$_4$), owing to the narrow band-width of the conduction (valence) band that helps to cutoff the thermal tail of the Boltzmann distribution at an energy window above (below) the conduction (valence) band. The thermal tail cutoff suppresses the thermionic leakage current in the OFF-state, thereby boosting the $I_\mathrm{ON}$/$I_\mathrm{OFF}$ ratio of all investigated devices to be greater than 10$^{3}$.

\subsubsection{Strain engineering} 

In-plane biaxial strain has been applied to tune the transport properties of \emph{p}-type doped $\alpha_{1(2)}$-MoSi$_2$N$_4$ and $\alpha_{1(2)}$-WSi$_2$N$_4$ DG FET that uses HfO$_2$ as the dielectric \cite{ghobadi2022_StrainedFET}. Compressive strains and tensile strains are used to tune the valleys at the valence bands of the monolayers, changing the effective mass of the holes and therefore the $I_\mathrm{ON}$/$I_\mathrm{OFF}$ ratios. $\Gamma$-valley is dominant over tensile strains while the K-valley increases for small compressive strain and then decreases for larger compressive strains. All the FETs display $I_\mathrm{ON}$/$I_\mathrm{OFF}$ ratios greater than 10$^{6}$. The ON-state current, OFF-State current and $I_\mathrm{ON}$/$I_\mathrm{OFF}$ ratio for all strained FETs are in the range of 2000–2200 $\mu$A $\mu$m$^{-1}$, 10$^{-3}$ $\mu$A $\mu$m$^{-1}$, and 2.0-2.2 $\times$ 10$^{6}$, respectively, whereas the $SS$ values lie in the range of 96–98 mV dec$^{-1}$.

\subsection{\label{sec: Contact_MA2Z4} NEGF Simulations of Electrical Contacts to MA$_2$Z$_4$}
Degenerately doping the source/drain is a common strategy employed in NEGF transport simulations to study material characteristics. However, in practicality, high doping is a difficult strategy to achieve in 2D semiconductors because of their ultra-thin nature \cite{zhang2019_2Dmaterials_doping}. Heavy doping can create bond dislocations and unwanted strains, leading to disruptive influences to the electronic properties of the contacted 2D semiconductor \cite{zhao2017_doping_TMDC}. Additionally, excessive localized charges introduced by the dopants can increase the rate of impurity scattering which leads to severe alteration of carrier mobility \cite{lee2023_localised_strain_doping}. Thus, directly contacting 2D semiconductor with external metallic electrodes remains the more practical approach for constructing 2D semiconductor FETs. 

NEGF simulations of metal contacts have been widely performed to obtain the LDDOS, which enables the transport gap - the energy barrier between the contacted region and the 2D semiconducting channel, to be extracted when the channel is under different $V_\mathrm{G}$. Below we provide a review on the NEGF simulations of metal/MoSi$_2$N$_4$ contacts under sub-10 nm device configurations. For DFT simulations of metal/MA$_2$Z$_4$ contacts, the readers are advised to consult a recent review \cite{Tho2023_MA2Z4_Review} that focuses on MA$_2$Z$_4$ contact heterostructures.

\subsubsection{3D metal contacts to MoSi$_2$N$_4$}

The transport properties of 3D metals contacts to MoSi$_2$N$_4$ FETs have been investigated using DFT-NEGF simulations \cite{Ying2024_MoSi2N4_FET, Zhanhai2024_MoSi2N4_FET}. Sc and Ti form \emph{n}-type Ohmic contact with MoSi$_2$N$_4$ at both the MS interface and the horizontal interface, which is in alignment with earlier DFT calculations \cite{Wang2021}. Furthermore, the metal/semiconductor (MS) contact interface formed by contacting (Sc, Ti)/MoSi$_2$N$_4$ exhibits zero vdW tunneling barrier, thus ensuring zero tunneling resistance across the vacuum gap. 

\begin{figure*}
\centering
\includegraphics[width=\textwidth]{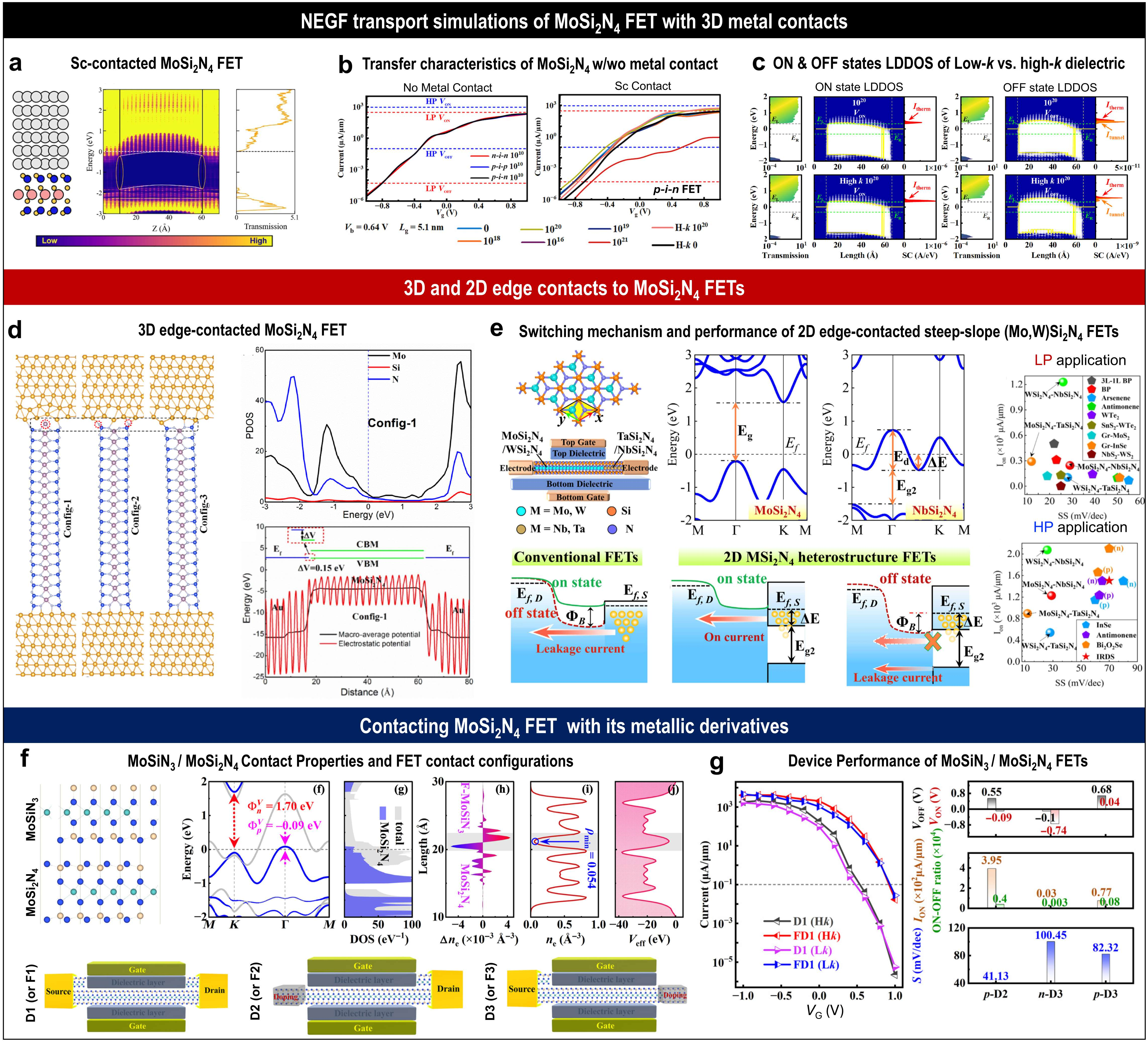}
\caption{\label{Fig10}\textbf{Quantum transport simulations of metal-contacted MA$_2$Z$_4$ FETs.} (a) LDDOS and transmission spectra of MoSi$_2$N$_4$ FET using Sc electrodes \cite{Ying2024_MoSi2N4_FET}. (b) Transfer characteristic of MoSi$_2$N$_4$ FET with source/drain doping, and with Sc electrodes alongside the use of different source/drain doping concentrations and low/high-$\kappa$ dielectric \cite{Zhanhai2024_MoSi2N4_FET}. (c) LDDOS and transmission spectra of MoSi$_2$N$_4$ FET at the ON and OFF states, using low/high-$\kappa$ dielectric and Sc as the electrode \cite{Zhanhai2024_MoSi2N4_FET}. (d) Different edge contact configurations of MoSi$_2$N$_4$ with Au as edge electrodes \cite{Meng2022_Au}. Also shown are the atom projected DOS, electrostatic potential profile and band edge alignment of one possible edge contact configuration. (e) Schematic of edge contact MSi$_2$N$_4$ (M = Nb, Ta, Mo, W) DG steep-slope FET, electronic band structures of MoSi$_2$N$_4$ and NbSi$_2$N$_4$ monolayers, and the leakage current mechanism in conventional FET versus that in MSi$_2$N$_4$ FET \cite{qu2023_MoSi2N4FET}. Comparisons of $I_\mathrm{ON}$ and $SS$ of MSi$_2$N$_4$ steep-slope FET with other 2D materials for HP and LP applications \cite{qu2023_MoSi2N4FET} are also shown. (f) Lattice structure of MoSi$_2$N$_3$/MoSi$_2$N$_4$ metal/semiconductor contact and the electronic properties (i.e. electronic band structure, density of states, plane-averaged charge density difference, valence electron density, effective potential) of MoSiN$_3$/MoSi$_2$N$_4$ \cite{li2025_MoSi2N4DerivativesFET}. Three different DG FET architectures are considered. (g) Transfer characteristic and performance metrics of the three different architectures with low/high-$\kappa$ dielectric \cite{li2025_MoSi2N4DerivativesFET}. (a) Copyright 2024, American Chemical Society; (b),(c) Copyright 2024, American Physical Society; (d) Copyright 2022, Elsevier B.V.; (e) Copyright 2023, IEEE; (f),(g) Copyright 2025, American Physical Society.}
\end{figure*}

In Fig. \ref{Fig10}(a), the LDDOS and the transmission spectra of MoSi$_2$N$_4$ with Sc are shown \cite{Ying2024_MoSi2N4_FET}. 
The LDDOS profiles show that Sc (also In and Ti) forms Ohmic contact with MoSi$_2$N$_4$ at the horizontal interface, with significant band bending in the channel for the FET formed with Sc (also Ti) contact. This horizontal band bending occurs when charges are redistributed at the horizontal interface between the MS contact and the MoSi$_{2}$N$_{4}$ channel, indicating the formation of covalent-liked bonding that also drastically modifies the electronic properties of MoSi$_{2}$N$_{4}$ beneath the metal. The Fermi level pinning factor, which quantifies the extent to which $E_F$ at MS interface deviates from the ideal Schottky-Mott limit, is found to be 0.52 and 0.53 for electrons and holes at the vertical interface, while 0.51 for both carriers at the horizontal interface \cite{Ying2024_MoSi2N4_FET}. These values are significantly higher than those of typical 2D semiconductors, suggesting superior tunability of the SBH when contacting 3D metals with MoSi$_2$N$_4$. 

The performance of MoSi$_2$N$_4$ FET with $L_\mathrm{G}$ = 5.1 nm using Sc and Cr contacts have been benchmarked against ITRS 2013 standards for HP and LP applications \cite{Zhanhai2024_MoSi2N4_FET}. Three types of DG FETs (\emph{n-i-n}, \emph{p-i-p}, and \emph{p-i-n}) with SiO$_2$ as the dielectric are first considered without contacting with the metals. All three types of FETs exhibit the same transfer characteristic under similar doping concentrations and device parameters [Fig. \ref{Fig10}(b)] and only \emph{p-i-n} FET was considered for further analysis. Even when Sc/Cr contacts are included, with various doping concentrations added to the source/drain of the \emph{p-i-n} FET, none of the devices can meet the $I_\mathrm{ON}$ requirements for both HP and LP applications [Fig. \ref{Fig10}(b)], although the use of HfO$_2$ as high-$\kappa$ dielectric with higher doping concentration do increase $I_\mathrm{ON}$. The LDDOS and transmission spectra profiles of Sc contacted \emph{p-i-n} device is shown in Fig. \ref{Fig10}(c), showing the change of the transport barrier when the device is switched between the ON-and OFF-state. For the Sc contacted device that is \emph{n}-type doped with a concentration of 10$^{20}$ cm$^{-3}$ at the source/drain, the transmission spectra current reveals that HfO$_2$ dielectric increases both the thermionic and tunneling current of the device at the ON/OFF-state. This higher spectra current when HfO$_2$ is used, elucidates the reason for the ON-state current being higher by an order of magnitude than the device that uses SiO$_2$.   

Motivated by the large design space of stacking 2D materials to form vdW MS contact, nine hundred types of Au/MoSi$_2$N$_4$ MS contact stacking have been screened based on their binding energies, whereby six configurations were selected for analysis \cite{Meng2022_Au}. 
Using PBE and HSE (Heyd–Scuseria–Ernzerhof) calculations, six MS contact configurations are found to form \emph{n}-type Schottky contact at the vertical interface between metal and semiconductor. The zero-bias transmission spectrum of these six MS contact configurations shows \emph{n}-type transport gap, in the range of 0.42-0.62 eV. Laterally stitched Au electrode and MoSi$_2$N$_4$ channel are also investigated as an edge-contacted device [Fig. \ref{Fig10}(d)]. The edge contacts are found to form \emph{p}-type Schottky contact at the lateral interface, with a lateral SBH = 0.15 eV lower than those of the vertical Au contacts, thus suggesting the edge contact configuration as a feasible route to improve the contact quality of MoSi$_2$N$_4$ FET.

\subsubsection{2D metal contacts to MoSi$_2$N$_4$} 

Laterally stitched (Mo, W)Si$_2$N$_4$ to (Nb, Ta)Si$_2$N$_4$ MS heterostructure have been investigated as a DG FET and diode, modeled with a channel length of 12 nm without any underlap \cite{qu2023_MoSi2N4FET}. The $E_F$ of monolayer (Nb, Ta)Si$_2$N$_4$ cuts across an isolated energy band, enabling the cut-off of the Fermi-Dirac distribution tail at the energy gaps above and below the isolated energy band, leading to the $SS$ achieving sub-thermionic limit [Fig. \ref{Fig10}(e)]. Notably, sub-thermionic $SS$ below 60 mV dec$^{-1}$ are obtained, with MoSi$_2$N$_4$/TaSi$_2$N$_4$ reaching 27 mV dec$^{-1}$. Benchmarking against IRDS 2020 standard for LP applications, the $I_\mathrm{ON}$ of the FETs exceeds 10$^{2}$ $\mu$A $\mu$m$^{-1}$ (e.g. $I_\mathrm{ON}$ of WSi$_2$N$_4$/NbSi$_2$N$_4$ reaches 1210 $\mu$A $\mu$m$^{-1}$ and $SS$ reaches 28 mV dec$^{-1}$), which is superior than the many other 2D semiconductor FETs \cite{liu2018_diracelectrons, lyu2020_SteepSlope, li2019_negativecapacitanceFET, chen2016_phosphoreneFET, szabo2015_MoTe2SnS2FET}. For HP applications, the $I_\mathrm{ON}$ of MoSi$_2$N$_4$/NbSi$_2$N$_4$ and WSi$_2$N$_4$/NbSi$_2$N$_4$ can reach 1240 $\mu$m$^{-1}$ and 2060 $\mu$m$^{-1}$, respectively, which fulfills the IRDS 2020 requirement. 

Metallic derivatives of MoSi$_2$N$_4$ have been studied as electrical contacts, such as MoSiN$_3$ using DFT and NEGF simulations \cite{li2025_MoSi2N4DerivativesFET}. The MoSiN$_3$ exhibits a Janus morphology and can form $p$-type contact with nearly Ohmic properties with MoSi$_2$N$_4$ when an appropriate contact face is used (denoted as F-MoSiN$_3$/MoSi$_2$N$_4$) [Fig. \ref{Fig10}(f)]. Three different FET architectures have been investigated [Fig. \ref{Fig10}(f)]. The first type, D1 (or FD1), employs MoSiN$_3$/MoSi$_2$N$_4$ (or F-MoSiN$_3$/MoSi$_2$N$_4$) vdW MS contact for both source and drain. The second type, D2 (or FD2), uses doped MoSi$_2$N$_4$ as the source electrode, with MoSiN$_3$/MoSi$_2$N$_4$ (or F-MoSiN$_3$/MoSi$_2$N$_4$) vdW MS contact as the drain. The third type, D3 (or FD3), inverts the D2 (or FD2) configuration, uses MoSiN$_3$/MoSi$_2$N$_4$ (or F-MoSiN$_3$/MoSi$_2$N$_4$) MS vdW contact as the source while using doped MoSi$_2$N$_4$ as the drain. At the horizontal interface, D1 and FD1 are found to exhibit ultralow \emph{p}-type SBH and \emph{p}-type quasi-Ohmic contact, respectively. When HfO$_2$ is used as the high-$\kappa$ dielectric for D1 and FD1, the $I_\mathrm{ON}$ are drastically improved [Fig. \ref{Fig10}(g)], and the addition of UL = 1.5 and 2.0 nm to FD1 allows the FET to achieve sub-thermionic $SS$. As for the two other types of architectures, \emph{p}-doped D2 device of doping concentration at 10$^{20}$ cm$^{-3}$ is found to achieve the lowest $SS$ value of 41.13 mV dec$^{-1}$ [Fig. \ref{Fig10}(g)]. It should be noted that none of the proposed architecture fulfills the $I_\mathrm{ON}$ lower limit of ITRS 2013 HP applications, suggesting that more efforts are needed to identify other metallic etched derivatives of MA$_2$Z$_4$ for high-performance FET application.

\subsubsection{3D metal contacts to CrX$_2$N$_4$ (X = C, Si)}

The contact properties of CrX$_2$N$_4$ (X = C, Si) FETs have been investigated using NEGF methods, by contacting the semiconductor with Ag, Au, Cu, Ni, Pd, Pt, Ti, and graphene as metal electrodes without applying any voltage bias \cite{shu2023_CrX2N4FET}. CrC$_2$N$_4$ forms an \emph{n}-type Ohmic contact with Ti electrodes at the vertical contact interface, a small \emph{n}-type SBH at the lateral interface, and an absence of a tunneling barrier at the vdW gap, which together suggest efficient electron injection. For CrSi$_2$N$_4$, the metals Ag, Au, Ni, Pd, Pt, and Ti form \emph{n}-type Ohmic contacts at the vertical contact interface, while Ag and Ti form \emph{n}-type Ohmic contacts at the lateral interface. Interestingly, band hybridization is almost absent in all these CrX$_2$N$_4$ MS contacts, such that the electronic band structure of the semiconductor underneath the contacting metal is well preserved, as compared to the case of Ti/MoSi$_2$N$_4$ \cite{Ying2024_MoSi2N4_FET} described earlier, whereby band hybridization is prevalent. The tunneling efficiency of the MS contacts can be quantified using the tunneling probability and tunneling specific resistivity. High values of the tunneling probabilities of 100\% have been found for Ti/CrC$_2$N$_4$, 50.48\% for Cu/CrC$_2$N$_4$ and 88.74\% for Ni/CrC$_2$N$_4$, compared to the other CrC$_2$N$_4$ MS contacts which show significantly lower tunneling probabilities. For the tunneling resistivity, the calculated values found in both CrX$_2$N$_4$ MS contact lie in the range of 0.005-0.091$\times$10$^{-9}$ $\Omega$ cm$^2$, which are lower than the values found in metal/(Mo, W)Si$_2$N$_4$ MS contacts \cite{Wang2021}. Ti demonstrates the highest contact performance in both CrC$_2$N$_4$ and CrSi$_2$N$_4$ FETs, suggesting that its potential in forming low contact resistance that is beneficial for charge injection.

\section{\label{sec: opportunites_and_challenges}Challenges and future directions of MoSi$_2$N$_4$ family transistors}

\begin{figure*}
\centering
\includegraphics[width=\textwidth]{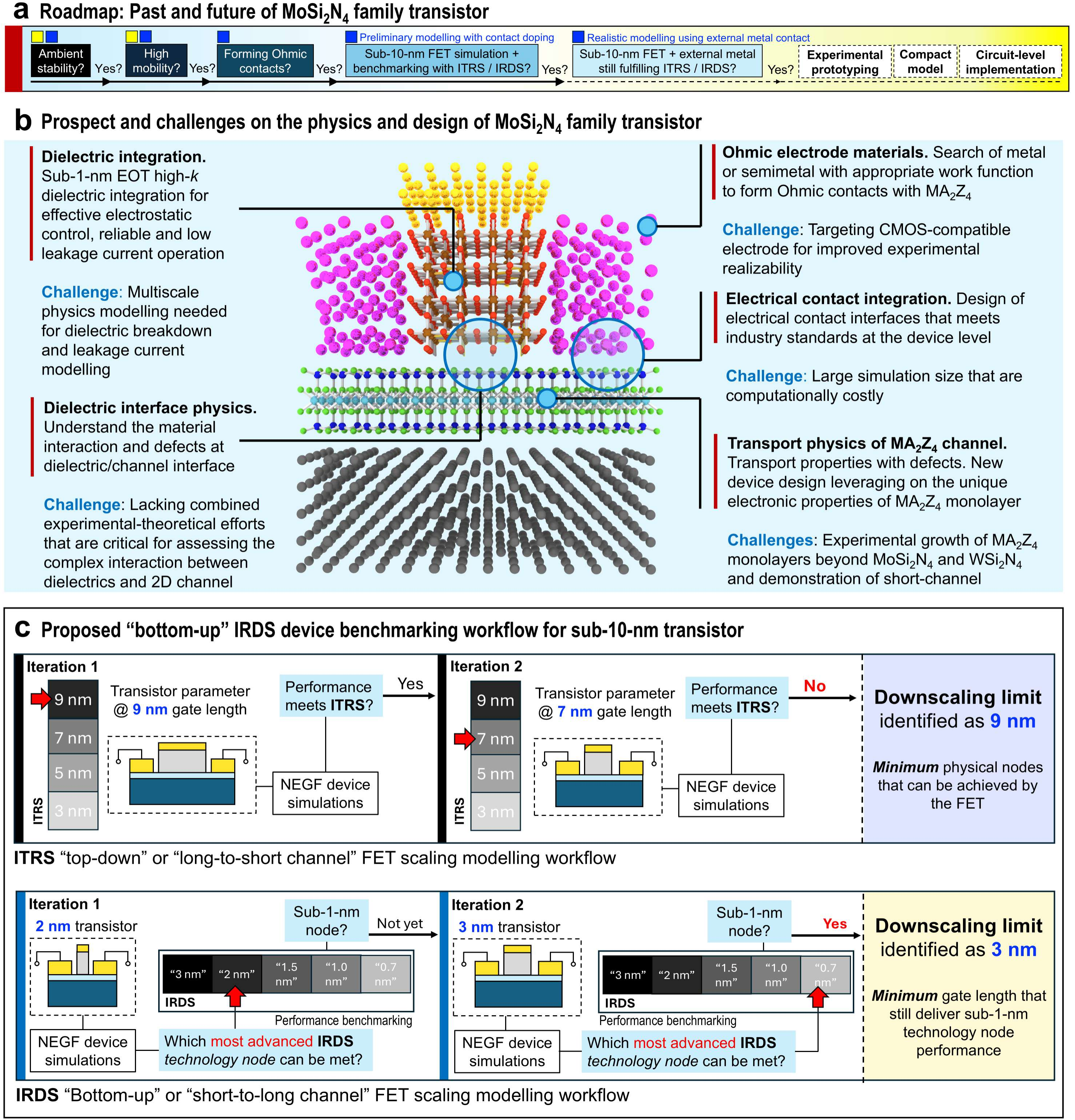}
\caption{\label{Fig11}\textbf{Challenges and future directions of MoSi$_2$N$_4$ family transistors.} (a) A simplified roadmap showing the past, now and future of MoSi$_2$N$_4$ monolayer transistor research. Device modelling that rigorously include external electrical contacts, experimental device prototyping, construction of compact models specific for MoSi$_2$N$_4$ family, and the compact integration of MA$_2$Z$_4$ FETs into circuits shall form the next research frontier. (b) Key challenges in MoSi$_2$N$_4$ family transistor simulation, including dielectric engineering, ohmic contact design, and transport modeling of MoSi$_2$N$_4$ channels, emphasizing the need for multiscale theoretical–experimental efforts. (c) Realigning the device simulation benchmarking workflow from ITRS2013 to IRDS. The ITRS device benchmarking adopts a `top-down' or `long-to-short channel' workflow, where $L_\mathrm{G}$ is shortened in each simulation iteration until the device can no longer meet the ITRS standards. The `last-pass' $L_\mathrm{G}$ is then quoted as the ultimate downscaling limit of the transistor. Benchmarking with IRDS will require a radically different, `bottom-up' or `short-to-long channel' approach, which starts with the most aggressively scaled $L_\mathrm{G}$. The $L_\mathrm{G}$ is then progressively increased in each simulation iteration until the first sub-1-nm technology node requirement indicated by IRDS is reached. The $L_\mathrm{G}$ is then quoted as the minimum gate length of a specific channel material in meeting the IRDS requriements.}
\end{figure*}

The ambient stability, carrier mobility higher than MoS$_2$ \cite{Hong2020}, compatibility of forming high-efficiency $n$-and $p$-type Ohmic contacts \cite{Wang2021}, and the excellent device-level performance predicted by NEGF simulations on sub-10-nm transistor setup \cite{hasani2023_FETMA2N4, Sun2021_transistor,Huang2021_transistor,Nandan2021_transistor,khorram2024_MoSi2N4FET,li2023_WSi2N4,Zhao2021_WGe2N4,Dong2023_MoSi2As4,Dong2023_WSi2P2As2,hasani2023_FETMA2N4,nandan2024transistors_MoSi2N4FET,nandan2023_EtchPassivationFET,ghobadi2022_StrainedFET}, provide a strong rationale in further exploring MoSi$_2$N$_4$ in sub-10-nm FET applications. More rigorous (and computationally more costly) device modeling that explicitly includes external metal contacts, the experimental prototyping of short-channel devices, and the circuit-level implementation \cite{huang20222d} such as compact model constructions \cite{cao2014compact,suryavanshi2016s2ds} are expected to form the key objectives in the next-stage research of MA$_2$Z$_4$ transistor [Fig. \ref{Fig11}(a)], in addition to challenges like Defects \cite{ducry2024_defectFET}, designing unconventional device architectures such as tunnelling FET (TFET) \cite{Ren2022, kanungo2022_2DTFET, sasioglu2025_multifunctional,salami2022_MoS2/WTe2_TFET}, and the experimental obstacles in growing high-quality 2D materials \cite{xu2022_2Dwafergrowth, geng2018_2Dmaterials_growth, chiappe2018_2Dmaterials_growth, zhang2019_2Dmaterials_doping}. In terms of computational design method, emerging techniques such as machine learning \cite{tsang2025_multiscale_machine_learning_modeling, wang2020_machine_learning_material_scientists}, offer exciting possibilities to reduce the computational cost of NEGF simulations. Figure \ref{Fig11}(b) summarizes some of the challenges and prospects of MA$_2$Z$_4$ FETs. In the following, we outline several prospective directions of MoSi$_2$N$_4$ FET for further exploration.

\subsection{Direction 1: Aligning sub-10-nm FET simulations with IRDS}

Computational simulations of sub-10-nm transistors are commonly benchmarked against semiconductor industry roadmaps. Historically, this involved aligning with the ITRS using a `\textit{top-down}' simulation methodology. In this approach, transistor performance at progressively smaller $L_\mathrm{G}$ was calculated and benchmarked against the ITRS's explicit, quantitative targets for each technology node. This `long-to-short channel' progression identified the ultimate scaling limit for a given transistor design when it could no longer meet the ITRS requirements.

To align the computational simulations of novel channel materials such as 2D semiconductors with IRDS, we propose a distinct `\textit{bottom-up}' approach. This approach begins by simulating the transistor at its most aggressive physical gate length such as $L_\mathrm{G} = 2$ nm. The calculated performance metrics are then evaluated against IRDS performance targets to identify the most advanced technology node that can be met by the device.
Subsequently, $L_\mathrm{G}$ is progressively \textit{increased} until the device enters the first \AA ngstr\"om (or sub-1-nm) technology node. This `short-to-long channel' workflow identifies the smallest $L_\mathrm{G}$ needed to meet or surpass the IRDS's projections for \AA ngstr\"om-era nodes. The rationale for targeting the \AA ngstr\"om node is due to its tremendous challenges for traditional silicon CMOS, thus justifying the need for novel channel materials. Through this workflow, the identified ultimate physical gate length ($L_\mathrm{G}^{(\text{min})}$) at which the device starts to meet the first \AA ngstr\"om node performance targets will then serve as an indicator of its potential for driving CMOS technology into, or even beyond, the \AA ngstr\"om era. Specifically, 2D semiconductor transistor for which the identified $L_\mathrm{G}^{(\text{min})}$ exceeds 10 nm would be deemed unviable, as they would not offer the necessary scaling benefits or competitive edge required to replace advanced silicon-based transistors in the sub-10-nm domain.

\textbf{Direction 2: Dielectric and electrical contact integrations.}
We propose that integrating the dielectric and electrical contacts is the next crucial step to more accurately assess the performance and optimize the design of MA$_2$Z$_4$ FETs. Dielectric integration is critical for 2D FET device performance \cite{venkatakrishnarao2024_liquid_metaloxide_highk_dielectrics, lau2023_dielectrics_TMD}. In NEGF-based quantum atomistic device simulations, the gate dielectric is typically modeled as an idealized dielectric environment. While this approach allows for computationally manageable simulations at the device level, it fundamentally neglects the atomistic interfacial interactions between the gate dielectrics and the 2D channel. As a result, important phenomena like band alignment \cite{jiang2023_lanthanum_oxyhalide}, electrical stability \cite{knobloch2022_amorphousgateoxide_Femileveltuning}, interface trap effects \cite{ghosh2025_hysteresis_MoS2_FET}, and leakage current \cite{knobloch2021_hBN_CMOS} cannot be fully captured. We propose that in-depth first-principles simulations should be conducted to understand the interface and device physics of integrating dielectrics with MA$_2$Z$_4$.

NEGF-based device simulations of MA$_2$ZZ$_4$ FETs currently simplify the electrical contact by arbitrarily adopting an optimal doping concentration at the source/drain regions. Unlike silicon, which can be heavily doped, doping 2D materials to form metallic contacts is challenging due to their atomically thin nature \cite{zhao2017_doping_TMDC, luo2019_functionalization_TMC}. Therefore, the explicit inclusion of external metals as source/drain electrodes in device modeling \cite{zhao2022_MoS2_Mo_FET} is an essential next step in the computational design studies of MA$_2$Z$_4$. Prior NEGF simulations of Sc-and Cr-contacted MoSi$_2$N$_4$ FET shows that fails to meet the $I_\mathrm{ON}$ requirement of IRDS standards even when high source/drain doping level is arbitrarily introduced \cite{Zhanhai2024_MoSi2N4_FET}, thus suggesting the complication and challenges of actually integrating external metal contact in FET simulaton. The grand challenge here would thus be the identification of $n$-and $p$-type Ohmic contacts with low interfacial tunneling barrier and good lattice matching to MA$_2$Z$_4$, which will enable a more realistic assessment of the FETs while maintaining a tractable simulation size.

\subsection{Direction 3: Device architectures beyond FET} 

Beyond conventional FET, MA$_2$Z$_4$ can be used for spin-based computing electronics since many magnetic MA$_2$Z$_4$ members have been predicted to exhibit non-trivial magnetic electronic band structures, such as half-metal, half-semiconductor and bipolar-magnetic-semiconductor \cite{Ren2022}. Spin transistor and filter shall be another frontier to explore. Tunneling field-effect transistors (TFETs) based on 2D semiconductors exhibit sub-thermionic $SS$, due to the subthreshold current being dominated by quantum tunneling instead of thermionic injection \cite{kanungo2022_2DTFET}, which further supports the existence of negative differential resistance (NDR). For example, TFET based on MoS$_2$/h-BN/MoS$_2$ vertical tunneling mechanism \cite{srivastava2016_MoS2/hBN/MoS2_TFET} can achieve sub-thermionic $SS$ of 57 mV dec$^{-1}$ and NDR at several $V_\mathrm{ds}$, and WS$_2$/SnS$_2$ TFET is able to achieve steep-slope $SS$ and $I_\mathrm{ON}/I_\mathrm{OFF}$ ratio in the order of 10$^{6}$ \cite{wang2018_WS2/SnS2_TFET}. While tunneling current has been observed to surpass thermionic current in the OFF-state of various MA$_2$Z$_4$ FETs, none of the studied device architectures had been specifically designed to leverage on the tunneling mechanism. For MA$_2$Z$_4$, we anticipate MA$_2$Z$_4$ TFET based on lateral \cite{xiong2020transverse} or vertical \cite{sarkar2015subthermionic} tunneling architecture to provide an alternative route to achieve ultralow power devices beyond conventional FET.  
Furthermore, the resistive switching behaviors \cite{li2021anomalous,pam2022interface} and neuromorphic device integration \cite{shi2018electronic, jain2025heterogeneous} of MoSi$_2$N$_4$ family remains unexplored thus far. First-principles simulations \cite{villena2024density} and quantum atomistic device modeling \cite{mitra2024atomistic} shall unveil the potential of MoSi$_2$N$_4$ family in this emerging research direction, where 2D semiconductors have shown promising performance \cite{huh2020_memristors,zhou2024recent} as reconfigurable logics \cite{pan2020reconfigurable}, nonvolatile memory \cite{yang2022two} and neuromorphic devices such as spiking neuron \cite{huo2024compact}, atomristor \cite{ge2018atomristor, yuan2025chip} and intelligent machine vision \cite{yang2024sensor}.

\subsection{Direction 4: Homologous metal contacts}

The \emph{homologous} metallic counterparts of MoSi$_2$N$_4$, namely the ultrathick monolayer of MoSi$_2$N$_4$(MoN)$_{4n}$ \cite{NSR, tho2025_zerodipole_contact} offers a route for electrode integration in MoSi$_2$N$_4$ FET. Recent DFT simulation reveals the formation of zero-dipole Schottky contact in MoSi$_2$N$_4$(MoN)/MoSi$_2$N$_4$ metal/semiconductor heterostructure due to the nearly identical chemical composition at the contact interface. The `ideal' Schottky-Mott limit is surprisingly preserved even in the cases of strong interfacial interaction \cite{tho2025_zerodipole_contact}. The broader MA$_2$Z$_4$ family and their homologous counterparts remain largely unexplored and may offer another platform for enhancing the performance of MA$_2$Z$_4$ FET.

\subsection{Direction 5: MA$_2$Z$_4$ nanotube and transistor applications} 

The nanotube counterparts of 2D materials such as graphene \cite{franklin2013_Carbon_NanotubeFET, tans1998_roomtemperature_Carbon_NanotubeFET, yao2021_Carbon_NanotubeFET, allen2007_Carbon_NanotubeBiosensors} and TMDs like MoS$_2$ \cite{strojnik2014_MoS2_NanotubeFET} and WS$_2$ \cite{levi2013_WS2_NanotubeFET, tamersit2024_WS2_Nanotube_GasSensing} offer another material platform for transistor applications. The 1D structure of the nanotube allows device operation at low $V_\mathrm{G}$, fast switching, and high integration density \cite{yang2006_graphene_FET_advantage}, potentially driving transistor scaling towards the sub-1 nm technology nodes.
The nanotube counterparts of MoSi$_2$N$_4$, WSi$_2$N$_4$ and other members of the MA$_2$Z$_4$ family have yet to be studied, and we anticipate such MA$_2$Z$_4$ nanotubes to provide another futile ground for study at both the material and device level.

\subsection{Direction 6: Machine learning acceleration of computational device design}

ML has emerged as a powerful tool to accelerate device design and predict performance metrics such as carrier mobility, $SS$ and $I_\mathrm{ON}$/$I_\mathrm{OFF}$ ratios. By learning patterns from large datasets generated from ab initio simulations, experimental data, or multi-scale modeling pipelines, ML models can offer insights into complex systems that would be unfeasible with quantum transport simulations \cite{wang2020_machine_learning_material_scientists}. Recent studies have shown that ML approaches such as random forests, support vector machines, neural networks, and Gaussian process regression can predict key properties of FETs with high accuracy, especially when trained on features like material composition, structural parameters, or even electronic descriptors extracted from prior DFT results \cite{tsang2025_multiscale_machine_learning_modeling}. 
Recent demonstration of graph neural network for high-accuracy simulation of \textit{pn} junction and FET without costly NEGF simulations reveals an exciting route to accelerate the modelling 2D channel FETs \cite{xie2025bottom}. As defects introduced during the synthesis process of 2D FET \cite{rai2022_defect_devices,knobloch2023_modeling_defect} can significantly complicate their design, ML-based approach could be used to correlate processing parameters with device performance, or to identify optimal material combinations and geometries for enhanced operation. ML can also be used to accelerate device characterization, such as the extraction of threhold voltage \cite{choi2023_ML_FET_VT}, which shall offer a powerful tool in the experimental device research stage of 2D channel FETs composed of MoSi$_2$N$_4$ family and the broader 2D semiconductor family. 

\section{\label{sec: conclusion}Conclusion}

As the semiconductor industry progresses into the \AA ngstr\"om-era technology nodes, silicon-based transistors are increasingly constrained by fundamental limitations, including severe short-channel effects, mobility degradation, elevated static power consumption, and the growing complexity of device fabrication processes \cite{akinwande2019graphene}. In this review, we consolidated the current state of research on the MA$_2$Z$_4$ family of two-dimensional materials—particularly MoSi$_2$N$_4$—as promising alternative channel candidates for sub-10 nm field-effect transistors. Notably, NEGF simulations have emerged as an indispensable tool for predicting, analyzing, and optimizing the quantum transport characteristics of MoSi$_2$N$_4$ transistors at these aggressive scaling limits.

Despite their compelling theoretical performance, significant challenges persist for the broader adoption of MoSi$_2$N$_4$ family for device technology. These include the experimental synthesis of defect-free, high quality monolayers, the scalable fabrication of ultra-short-channel devices, and the development of computational models that accurately captures realistic device conditions such as electrical contacts, dielectric interfaces, and material defects. Addressing these challenges is not only critical for practical device integration but also offers a unique opportunity to deepen our understanding of the physics in 2D semiconductors.

By crystallizing recent advances in the computational design of sub-10-nm field-effect transistors, this review serves as a bridge between fundamental research and technological implementation. We anticipate that MoSi$_2$N$_4$ will continue to provide fertile ground for pushing the performance boundary of transistor beyond the current semiconductor technology paradigm.

\section*{Acknowledgments}
This work is supported by the Singapore National Research Foundation (NRF) Frontier Competitive Research Programme (F-CRP) under the award number NRF-F-CRP-2024-0001. J.L. acknowledges the supports from National Natural Science Foundation of China (No. 12274002 and
91964101) and the Ministry of Science and Technology of
China (No. 2022YFA1203904). L.K.A. acknowledges the support from the Singapore A*STAR IRG (Grant No. M23M6c0102).

\bibliography{citation}
\end{document}